\documentstyle[12pt]{article}

\newcommand{\mathR}{{\rm I\! R}}     
\newcommand\mathC{\mkern3mu\raise2.5pt\hbox{$\scriptscriptstyle|$}
        {\mkern-6.2mu\rm C}}

\begin{document}

\hfill {IMPERIAL/TP/98--99/45}


\begin{center}
    {\large Spacetime and the Philosophical Challenge of Quantum
Gravity}\footnote{To appear in {\em Physics meets
Philosophy at the Planck Scale\/}, ed.\ C.~Callender and
N.~ Huggett, Cambridge University Press (2000).}
\end{center}

\begin{center}
        J.~Butterfield\footnote{email:
        jb56@cus.cam.ac.uk, jeremy.butterfield@all-souls.oxford.ac.uk}\\[5pt]
        All Souls' College\\
        Oxford OX1 4AL\\
\end{center}

\begin{center}
        and
\end{center}

\begin{center}
        C.J.~Isham\footnote{email: c.isham@ic.ac.uk}\\[5pt]
        The Blackett Laboratory\\
        Imperial College of Science, Technology \& Medicine\\
        South Kensington\\
        London SW7 2BZ\\
\end{center}

\begin{center} 17 March 1999 \end{center}

\begin{abstract}
We survey some philosophical aspects of the search for a
quantum theory of gravity, emphasising how quantum gravity
throws into doubt the treatment of spacetime common to the
two `ingredient theories' (quantum theory and general
relativity), as a 4-dimensional manifold equipped with a
Lorentzian metric. After an introduction (Section
\ref{Sec:Intro}), we briefly review the conceptual problems
of the ingredient theories (Section \ref{Sec:ConceptProb})
and introduce the enterprise of quantum gravity (Section
\ref{Sec:WhatisQG}). We then describe how three main
research programmes in quantum gravity treat four topics of
particular importance: the scope of standard quantum
theory; the nature of spacetime; spacetime diffeomorphisms,
and the so-called problem of time (Section
\ref{Sec:ResearchProg}). By and large, these programmes
accept most of the ingredient theories' treatment of
spacetime, albeit with a metric with some type of quantum
nature; but they also suggest that the treatment has
fundamental limitations. This prompts the idea of going
further: either by quantizing structures other than the
metric, such as the topology; or by regarding such
structures as phenomenological. We discuss this in Section
\ref{Sec:TowardsQuST}.
\end{abstract}

\section{Introduction}\label{Sec:Intro}
\subsection{Prologue}\label{SubSec:Prologue}
Any branch of physics will pose various philosophical
questions: for example about its concepts and general
framework, and the comparison of these with analogous
structures in other branches of physics. Indeed, a
thoughtful consideration of any field of science leads
naturally to questions within the philosophy of science.

However, in the case of quantum gravity we rapidly
encounter fundamental issues that go well beyond questions
within the philosophy of science in general. To explain
this point, we should first note that by `quantum gravity'
we mean any approach to the problem of combining (or in
some way `reconciling') quantum theory with general
relativity.\footnote{Though this construal of `quantum
gravity' is broad, we take it to exclude studies of a
quantum field propagating in a spacetime manifold equipped
with a fixed background Lorentzian metric. That is to say,
`quantum gravity' must involve in some way a quantum
interaction of the gravitational field with itself. Quantum
field theory in a fixed background is a rich subject in its
own right, and it is a useful way of probing certain
aspects of quantum gravity. But it does not raise the
philosophical problems that we will pursue in this paper.}
An immense amount of effort has been devoted in the last
forty years to combining these two pillars of modern
physics. Yet although a great deal has been learned in the
course of this endeavour, there is still no satisfactory
theory: rather, there are several competing approaches,
each of which faces severe problems, both technical and
conceptual.

This situation means that there are three broad ways in
which quantum gravity raises questions beyond the
philosophy of science in general.
\begin{enumerate}
\item First, each of the `ingredient theories'---quantum theory
and general relativity---poses significant conceptual
problems in its own right. Since several of these problems
are relevant for various topics in quantum gravity (as Part
3 of this volume bears witness), we must discuss them,
albeit briefly. We will do this in Section
\ref{Sec:ConceptProb}, to help set the stage for our study
of quantum gravity. But since these problems are familiar
from the literature in the philosophy, and foundations, of
physics, we shall be as brief as possible.

\item Second, the fundamentally disparate bases of the
two ingredient theories generate major new problems when
any attempt is made to combine them. This will be this
paper's main focus. But even to summarize these new
conceptual problems is a complicated and controversial
task: complicated because these problems are closely
related to one another, and to the technical problems; and
controversial because what the problems are, and how they
are related to one another, depends in part on problematic
matters in the interpretation of the ingredient theories.
Accordingly, the main task of this paper will be just to
give such a summary; or rather, part of such a
summary---roughly speaking, the part that relates to the
treatment of spacetime.\footnote{We choose this topic
partly in the light of this volume's emphasis on spacetime,
especially in Parts 2 and 5.} We undertake this task in
Sections \ref{Sec:WhatisQG} et seq.

\item Third, the contrast between our lacking
a satisfactory theory of quantum gravity and our having
supremely successful ingredient theories, raises questions
about the nature and function of philosophical discussion
of quantum gravity. It clearly cannot `take the theory as
given' in the way that most philosophy of physics does; so
how should it proceed? Though we are cautious about the
value of pursuing meta-philosophical questions, we think
this question deserves to be addressed. We do so in the
course of Section \ref{SubSec:NoData}; (where we also
emphasise how unusual quantum gravity is, as a branch of
physics).
\end{enumerate}

So much by way of gesturing at the entire scope of
`philosophy of quantum gravity'. In the rest of this
Subsection, we shall make some brief general comments about
the source of the conceptual problems of quantum gravity,
and thereby lead up to a more detailed prospectus.

Despite the variety of programmes, and of controversies, in
quantum gravity, most workers would agree on the following,
admittedly very general, diagnosis of what is at the root
of most of the conceptual problems of quantum gravity.
Namely: general relativity is not just a theory of the
gravitational field---in an appropriate sense, it is also a
theory of spacetime itself; and hence a theory of quantum
gravity must have something to say about the quantum nature
of space and time. But though the phrase `the quantum
nature of space and time' is portentous, it is also very
obscure, and opens up a Pandora's box of challenging
notions.

To disentangle these notions, the first thing to stress is
that---despite the portentous phrase---there is, in fact, a
great deal {\em in common\/} between the treatments of
space and time given by the ingredient theories, quantum
theory and general relativity. Specifically: they both
treat space and time as aspects of spacetime, which is
represented as a $4$-dimensional differentiable manifold,
while the metrical structure of spacetime is represented by
a Lorentzian metric on this manifold.

In view of this, one naturally expects that a theory of
quantum gravity will itself adopt this common treatment, or
at least its main ingredient, the manifold conception of
space and time. Indeed, we will see in Section 4 that the
three main research programmes in quantum gravity do accept
this conception. So returning to our phrase, `the quantum
nature of space and time': although `quantum' might suggest
`discrete', the `quantum nature of space and time' need not
mean abandoning a manifold conception of space and time at
the most fundamental level. And consequently: the clash
mentioned above---between the disparate bases of the
ingredient theories, quantum theory and general
relativity---need not be so straightforward as the
contradiction between discreteness and continuity. As we
shall see in Sections \ref{Sec:WhatisQG} et seq., the clash
is in fact both subtle and multi-faceted.

But we will also see in Section \ref{Sec:ResearchProg} that
in various ways, and for various reasons, these programmes
do not accept all of the common treatment, especially as
regards the dimensionality and metric structure of
spacetime: the main difference being their use of some type of
quantized metric. Furthermore, the two main {\em current\/}
programmes even suggest, in various ways, that the manifold
conception of spacetime is inapplicable on the minuscule
length-scales characteristic of quantum gravity.

So the situation is curious: although the ingredient theories
have much in common in their treatments of space and time, this
common treatment is threatened by their attempted unification.
This situation prompts the idea of departing more radically from
the common treatment. Besides, quite apart from the challenges to
the manifold conception that come from the programmes in Section
\ref{Sec:ResearchProg}, other quantum gravity programmes reject
this conception from the outset. We explore these ideas in
Section \ref{Sec:TowardsQuST}.

Here it is helpful to distinguish two general strategies
for going beyond the common treatment, the first more
specific than the second. (1) One can quantize a classical
structure which is part of that treatment, and then recover
it as some sort of classical limit of the ensuing quantum
theory. (2) More generally, one can regard such a structure
as phenomenological, in the physicists' sense of being an
approximation, valid only in regimes where quantum gravity
effects can be neglected, to some other theory (not
necessarily a quantum theory). In more philosophical
jargon, this second strategy is to regard the classical
structure as emergent from the other theory---though here
`emergence' must of course be understood as a relation
between theories, not as a temporal process. As we will
see, both these strategies can be applied either to
metrical structure (as in the programmes in Section
\ref{Sec:ResearchProg}), or to structures required just by
the manifold conception, such as topology (Section
\ref{Sec:TowardsQuST}).

But whichever of these two strategies one adopts, there are
clearly two rather different ways of thinking about the
relation between the familiar treatment of space and
time---in common between the ingredient theories---and the
treatment given by the, as yet unknown, theory of quantum
gravity (with or without a manifold at the fundamental
level). First, one can emphasise the emergence of the
familiar treatment: its being `good enough' in certain
regimes. We adopt this perspective in a complementary
essay.\footnote{See \cite{BI99a}. It discusses
philosophical aspects of emergence, which here we only
treat very briefly, in Section \ref{Sec:TowardsQuST}. But
in another regard, it is more limited: it discusses the
emergence of spatio-temporal concepts for only one of the
three programmes in Section \ref{Sec:ResearchProg}.}
Second, one can emphasise instead `the error of its ways'.
That is, one can emphasise how quantum gravity suggests
limitations of the familiar treatment; (though the
conclusions of the examination will of course be tentative,
since we have no satisfactory theory of quantum gravity).
That is the perspective of this paper.\footnote{We should
add that though this paper focusses on the treatment of
spacetime, quantum gravity also `puts pressure' on the
formalism of, and usual interpretative ideas about, quantum
theory; as we shall briefly discuss in Section
\ref{Sec:ConceptProb}. Furthermore, quantum gravity even
puts pressure on standard mathematics itself, in that
constructing a theory of quantum gravity might require some
non-standard mathematical ideas: for example, spacetime
might be represented by something that is not a {\em
set\/}. We will briefly discuss this line of thought in
Section \ref{Sec:TowardsQuST}.}

In more detail, our plan is as follows. First we make some
general orienting remarks about philosophy of quantum
gravity (Section \ref{SubSec:NoData}) and realism (Section
\ref{SubSec:Realism?}). Then we briefly review the bearing
on quantum gravity of the conceptual problems of the
ingredient theories: quantum theory (Section
\ref{SubSec:InterpretQT}), and more briefly, general
relativity (Section \ref{SubSec:InterpretGR}). In Section
\ref{Sec:WhatisQG}, we introduce the enterprise of quantum
gravity proper. We state some of the main approaches to the
subject; summarise some of the main motivations for
studying it (Section \ref{SubSec:ApproachesQG}); and
introduce some of the conceptual aspects that relate
closely to spacetime: the role of diffeomorphisms (Section
\ref{SubSec:RoleDiffeos}) and the problem of time (Section
\ref{SubSec:ProblemTime}).

In Section \ref{Sec:ResearchProg}, we first set up the four
topics in terms of which we will survey three
well-developed research programmes in quantum gravity. The
topics are: (i) the extent to which the programme uses
standard quantum theory; (ii) the extent to which it uses
standard spacetime concepts; (iii) how it treats spacetime
diffeomorphisms; and (iv) how it treats the problem of
time. After some historical remarks (Section
\ref{SubSec:History}), we turn to our three research
programmes: the old particle-physics approach (Section
\ref{SubSec:ParticleProg}), supergravity and superstrings
(Section \ref{SubSec:Superstrings}), and canonical quantum
gravity (Section \ref{SubSec:CanWQG}). (We also briefly
treat a distinctive version of the latter, the Euclidean
programme.) By and large, these three programmes are
`conservative' in the sense that although they in some way
quantize metric structure, they use standard quantum
theory, and they treat the spacetime manifold in a standard
way. But in various ways, they also suggest that this
treatment of spacetime is phenomenological. To mention two
obvious examples: in the superstring programme, there are
suggestions that many {\em different\/} manifolds play a
role. And in the main current version of canonical quantum
gravity (the Ashtekar programme), there are suggestions
that quantities such as area and volume are quantised; and
that the underlying structure of spacelike may be more like
a combinatorial network than a standard continuum manifold.

This situation prompts the idea of a quantum gravity programme
that {\em ab initio\/} goes beyond the standard treatments of
spacetime and/or quantum theory. So in Section
\ref{Sec:TowardsQuST}, we discuss some of these more
radical ideas: quantizing spacetime structures other than
the metric, or regarding such structures as
phenomenological---where this could involve
abandoning the manifold conception in favour of a variety
of novel mathematical structures.

Finally, a caveat about the scope of the paper. Even for the
well-established research programmes, we shall omit, or mention only
in passing, many of the more recent ideas. This is partly because of
lack of space, and partly because technical ideas need to acquire a
certain degree of maturity before philosophical reflection becomes
appropriate. Similarly, we will not mention many of the more
speculative ideas that have emerged over the years. Instead, we have
deliberately limited ourselves to some of the well-established ideas
in a few of the well-established research programmes, and to just a
few of the more speculative ideas.

\subsection{No Data, No Theory, No
Philosophy?}\label{SubSec:NoData} Before embarking on this
detailed analysis of the conceptual challenge of quantum
gravity, it is important to briefly consider the general
idea of philosophical discussion of quantum gravity: more
pointedly, to defend the idea, in the face of the fact that
there is far from being a universally agreed theory of
quantum gravity! So in this Subsection, we will point out
the peculiarity of quantum gravity in comparison with other
branches of theoretical physics, and discuss how this
affects the way in which one can write about the subject
from a philosophical perspective. This will lead to some
discussion of the philosophical positions we intend to
adopt, especially as regards realism.

The most obvious peculiarity is a dire lack of data. That
is, there are no phenomena that can be identified
unequivocally as the result of an interplay between general
relativity and quantum theory---a feature that arguably
challenges the right of quantum gravity to be considered as
a genuine branch of science at all!

This lack of obvious empirical data stems from a simple
dimensional argument that quantum gravity has a natural
length scale---the Planck length defined using dimensional
analysis as\footnote{Here $G$ is Newton's constant, $\hbar$
is Planck's constant, and $c$ is the speed of light.}
$L_P:=(G\hbar/c^3)^{1\over2}$---and this is extremely
small: namely $10^{-35}$m. By comparison: the diameters of
an atom, proton and quark are, respectively, about
$10^{-10}$m, $10^{-15}$m, and $10^{-18}$m. So the Planck
length is as many orders of magnitude from the (upper limit
for) the diameter of a quark, as that diameter is from our
familiar scale of a metre! The other so-called `Planck
scales', such as the Planck energy and the Planck time, are
equally extreme: the Planck energy $E_P$ has a value of
$10^{22}$MeV, which is far beyond the range of any
foreseeable laboratory-based experiments; and the Planck
time (defined as $T_P:=L_P/c$) has a value of about
$10^{-42}$ seconds.

These values suggest that the only physical regime where
effects of quantum gravity might be studied directly---in
the sense that {\em something\/} very specific can be
expected---is in the immediate post big-bang era of the
universe---which is not the easiest thing to probe
experimentally! This problem is compounded by the fact
that, modulo certain technical niceties, {\em any\/}
Lorentz-invariant theory of interacting spin-2 gravitons
with a conserved energy-momentum tensor will yield the same
low-energy scattering amplitudes as those obtained from a
perturbative expansion of the Einstein Lagrangian. Thus
different quantum gravity theories might only reveal their
differences empirically at very high energies.

This lack of data has implications for both physics and
philosophy. For physics, the main consequence is simply
that it becomes very difficult to build a theory: witness
the fact that a fully satisfactory quantum theory of
gravity remains elusive after forty years of intense
effort. Of course, a great deal has been learnt about the
features that such a theory might possess, albeit partly by
eliminating features that are now known not to work. A good
example is the perturbative unrenormalizability of the
particle-physics approach to quantum gravity (see Section
\ref{SubSec:ParticleProg}).

But there is more to the difficulty of theory-construction
than just the lack of data: and this `more' relates to
philosophy in two ways. Firstly, this difficulty is partly
due to conceptual problems which clearly bear on
philosophical discussions of concepts such as space, time
and matter. Here we have in mind both kinds of conceptual
problem, listed as 1.\ and 2.\ in Section
\ref{SubSec:Prologue}: those arising from the disparateness
of the bases of general relativity and quantum theory, and
also problems about each of these theories in themselves.
To take an obvious example of the latter, quantum gravity
is usually taken to include quantum cosmology; and here,
the idea of a `quantum state of the universe' immediately
confronts conceptual problems about quantum theory such as
the meaning of probability, and the interpretation of the
quantum state of a closed system---in this case, the
universe in its entirety.\footnote{There is another
connection with the conceptual problems about quantum
theory. Namely, the lack of data in quantum gravity
research is analogous to that which---until relatively
recently---faced research in conceptual problems about
quantum theory. In view of these connections, it is curious
that there has been so little interaction between the two
research communities.} And as we shall see in Section
\ref{SubSec:InterpretGR}, such examples do not only come
from quantum theory, with its notorious conceptual
problems. They also come from general relativity---whose
foundations are murkier than philosophers commonly take
them to be. To sum up: the difficulty of
theory-construction is partly due to conceptual problems
(and is thereby related to philosophy), not just lack of
data.

Secondly, theory-construction is difficult because there is not
even agreement on what {\em sorts\/} of data a quantum theory of
gravity would yield, if only we could get access to them! More
precisely, the dimensional argument discussed above suggests that
only phenomena at these very small distances, or high energies,
would exhibit quantum gravity effects; which implies that the
main application of the quantum theory of gravity may be to the
physics of the very early universe.  However, this argument rests
on the assumption that any physically measurable quantity can be
expressed as a power series in a number like $E/E_P$ (where $E$
is some characteristic energy scale for an experiment), so that
the quantity's predicted value is tiny in any experiment that
probes energy-scales that are much lower than $E_P$. But
experience elsewhere in quantum field theory suggests that
`non-perturbative effects' could also occur, in which the
predicted values of certain measurable quantities are not
analytic functions of the coupling constant, and this totally
changes the argument: for example, if $x$ is a very small number,
then $|\log x|$ is very large! But, in the case of quantum
gravity it is anybody's guess what {\em sort\/} of effects would
be exhibited---and therefore, what sort of predictions the
envisaged quantum theory of gravity is meant to give. For
example, for all we know, it might predict the masses of all the
elementary particles.

The most obvious consequence of such uncertainty is that it
makes for a sort of circularity. On the one hand, it is
ferociously difficult to find the theory without the help
of data, or even an agreed conception of the {\em sort\/}
of data that would be relevant. On the other hand, we can
only apply our present theories to (and get evidence from)
regimes well way from those determined by the Planck scale;
so we cannot judge what phenomena might be relevant to a
theory of quantum gravity, until we know what the theory
is.

In this predicament, theory construction inevitably becomes
much more strongly influenced by broad theoretical
considerations, than in mainstream areas of physics. More
precisely, it tends to be based on various {\em prima
facie\/} views about what the theory {\em should\/} look
like---these being grounded partly on the philosophical
prejudices of the researcher concerned, and partly on the
existence of mathematical techniques that have been
successful in what are deemed, perhaps erroneously, to be
closely related areas of theoretical physics, such as
non-abelian gauge theories. In such circumstances, the goal
of a research programme tends towards the construction of
abstract theoretical schemes that are compatible with some
preconceived conceptual framework, and are internally
consistent in a mathematical sense.

This situation does not just result in an extreme
`underdetermination of theory by data', in which many
theories or schemes, not just a unique one, are presented
for philosophical assessment. More problematically, it
tends to produce schemes based on a wide range of
philosophical motivations, which (since they are rarely
articulated) might be presumed to be unconscious
projections of the cthonic psyche of the individual
researcher---and might be dismissed as such! Indeed,
practitioners of a given research programme frequently have
difficulty in understanding, or ascribing validity to, what
members of a rival programme are trying to do. This is one
reason why it is important to uncover as many as possible
of the assumptions that lie behind each approach: one
person's `deep' problem may seem irrelevant to another,
simply because the starting positions are so different.

This situation also underlines the importance of trying to
find some area of physics in which any putative theory
could be tested directly. A particularly important question
in this context is whether the dimensional argument
discussed above can be overcome, {\em i.e.}, whether there
are measurable quantum gravity effects well below the
Planck scales; presumably arising from some sort of
non-perturbative effect. However, the existence of such
effects, and the kind of phenomena which they predict, are
themselves likely to be strongly theory-dependent.

It follows from all this that the subject of quantum
gravity does not present the philosopher with a
conceptually or methodologically unified branch of physics,
let alone a well-defined theory; but instead with a wide
and disparate range of approaches. We must turn now to
discussing what this situation implies about the scope and
possible topics of philosophical discussion.

As we conceive it, philosophy of physics is usually
concerned with either: (i) metaphysical/ontological issues,
such as the nature of space, or physical probability; or
(ii) epistemological/methodological issues, such as
underdetermination of theories, or scientific realism, with
special reference to physics. And in both areas (i) and
(ii), the discussion is usually held in focus by
restricting itself to a reasonably well-defined and
well-established physical theory; or at worst, to a small
and homogeneous set of reasonably well-defined rival
theories. Hence the image of philosophy of physics as a
Greek chorus, commenting {\em post facto\/} on the dramatic
action taking place on a well-defined stage. But, as we
have seen, this kind of role is not available in quantum
gravity, where there are no reasonably well-defined, let
alone well-established, theories. So should philosophical
discussion simply wait until the subject of quantum gravity
is much better established?

We think not. For there are at least three (albeit
overlapping) ways in which to pursue the subject
`philosophy of quantum gravity'; each of them valuable, and
each encompassing many possible projects. The first two are
straightforward in that all will agree that they are
coherent endeavours; the third is more problematic. This
paper will be an example of the second way. Broadly
speaking, they are, in increasing order of radicalism:
\begin{enumerate}
\item {First, one can undertake
the `normal' sort of philosophical analysis of some
sufficiently specific and well-defined piece of research in
quantum gravity; even though the price of its being
well-defined may be that it is too specific/narrow to
warrant the name `theory'---so that, in particular, it
would be unwise to give much credence to the ontological or
interpretative claims it suggests. An example might be one
of the specific perturbative approaches to string
theory---for example, the type II-B superstring---which
recent research in the area suggests might merely be one
particular perturbative regime, of an underlying theory
(`$M$-theory') whose mathematical and conceptual structures
are quite different from those of a quantised loop
propagating in a spacetime manifold. This sort of analysis
is certainly valuable, even if the limitations of any
specific example must make one wary of any ontological or
interpretative suggestions which arise. For examples of
such analyses, see the papers in Parts 3 and 4 of this
volume. }

\item {Second: one can try to relate a {\em range\/} of
conceptual problems about general relativity, quantum
theory, and their having disparate bases, to a {\em
range\/} of approaches or research programmes in quantum
gravity.

To make this endeavour manageable, one must inevitably
operate at a less detailed level than in 1.\ above. The
hope is that despite the loss of detail, there will be a
compensating value in seeing the overall pattern of
relationships between conceptual problems and
mathematical/physical approaches to quantum gravity.
Indeed, one can hope that such a pattern will be
illuminating, precisely because it is not tied to details
of some specific programme that may be on the proverbial
hiding to nowhere.

Such a pattern of relationships can be envisioned in two
ways, according to which `side' one thinks of as
constraining the other. One can think of problems and ideas
about general relativity and quantum theory as giving
constraints on---or heuristic guides to our search
for---quantum gravity. (Of course, we harbour no illusions
that such conceptual discussion is a {\em prerequisite\/}
for successful theory-construction---of that, only time
will tell.) And {\em vice versa\/}, one can think of
quantum gravity as constraining those problems and ideas,
and even as suggesting possible changes to the foundations
of these constituent theories. In what follows, we shall
see examples of both kinds of constraint.}

\item {Third, one can try to study quantum gravity
in the context of some traditional philosophical ideas that
have nothing to do with the interpretation of general
relativity and quantum theory {\em per se\/}---for example,
traditional concepts of substance and attribute. Again, one
can think of such a relation in two ways: the philosophical
idea giving constraints on quantum gravity; and {\em vice
versa\/}, quantum gravity reflecting back on the
philosophical idea.

We admit to finding this endeavour alluring. But, again, we
harbour no illusions that such traditional philosophical
ideas are {\em likely\/} to be heuristically helpful, let
alone a prerequisite, for theory-construction. Similarly,
one must be tentative about constraints in the opposite
direction; {\em i.e.\/} about the idea that a traditional philosophical
position could be `knocked out' by a quantum gravity
proposal, where `knocked out' means that the position is
shown to be, if not false, at least `merely'
phenomenological, or approximately true, in ways that
philosophers tend not to realize is on the cards. Since no
quantum gravity proposal is well-established, any such
`knock-out' is tentative.}
\end{enumerate}

This paper will exemplify type 2.\ above, although we
emphasise that we are sympathetic to type 3. More
generally, we are inclined to think that in the search for
a satisfactory theory of quantum gravity, a fundamental
reappraisal of our standard concepts of space, time and
matter may well be a necessary preliminary. Thus we are
sceptical of the widespread idea that at the present stage
of quantum gravity research, it is better to try first to
construct an internally consistent mathematical model and
only then to worry about what it `means'. But such a
reappraisal is fiercely hard to undertake; and accordingly,
this paper adopts `the middle way'---type 2.

A final remark about the various ways to pursue `philosophy
of quantum gravity'. Though type 2.\ involves, by
definition, surveying ideas and approaches, this by no
means implies that 2.\ encompasses a single project, or
even that it encompasses only one project focussing on the
nature of space and time. For example, here is an
alternative project, relating to traditional positions in
the philosophy of geometry.

The immense developments in pure and physical geometry from
Riemann's {\em habilitationschrift\/} of 1854 to the
establishment of general relativity, transformed the
philosophy of geometry beyond recognition. In particular,
Kant's apriorism fell by the wayside, to be replaced by
empiricism and conventionalism of various stripes. With
this transformation, the idea that at very small
length-scales, space might have a non-manifold structure
became a `live option' in a way that it could not have been
while Kant's influence held sway in its original form. Yet,
in fact, this idea has had only a small role in the
philosophy of geometry of the last 150 years---for the
perfectly good reason that no significant physical theory
took it up. (Its main role is via Riemann's view---endorsed
in our own day by Grunbaum---that in a discrete space, but
not a manifold, the metric is, or can be, intrinsic, and
thereby non-conventional.)

But nowadays, there are several `unconventional'
approaches to quantum gravity that postulate a non-manifold
structure for spacetime; and even in the more conventional
approaches, which do model space, or spacetime, with a
differentiable manifold, there are often hints of a
discrete structure that lies beneath the continuum picture
with which one starts. For example, in the Ashtekar
programme, area and volume variables become discrete; and
in superstring theory there are strong indications that
there is a minimal size for length.

These proposals prompt many questions for philosophers of
geometry; the obvious main one being, how well can the
traditional positions---the various versions of empiricism
and conventionalism---accommodate such proposals? We will
not take up such questions here; though we like to think
that this paper's survey of issues will help philosophers
to address them.

\subsection{Realism?}\label{SubSec:Realism?}
Finally, we should briefly discuss the bearing of our
discussion on the fundamental questions of realism. Thus it
is natural to ask us (as one might any authors in the
philosophy of physics):---Does the discussion count for or
against realism, in particular scientific realism; or does
it perhaps presuppose realism, or instead its falsity?

Our answer to this is broadly as follows. We will write as
if we take proposals in quantum gravity realistically, but
in fact our discussion will not count in favour of
scientific realism---nor indeed, against it. This lack of
commitment is hardly surprising, if only because, as
emphasised in Section \ref{SubSec:NoData}, quantum gravity
is too problematic as a scientific field, to be a reliable
test-bed for scientific realism. But we will fill out this
answer in the rest of this Subsection. In short, we will
claim: i) we are not committed to scientific realism; and
ii) there is a specific reason to be wary of reifying the
mathematical objects postulated by the mathematical models
of theoretical physics. We shall also make a comment
relating to transcendental idealism.

\subsubsection{Beware Scientific Realism}
Scientific realism says, roughly speaking, that the
theoretical claims of a successful, or a mature, scientific
theory are true or approximately true, in a correspondence
sense, of a reality independent of us. So it is a
conjunctive thesis, with an ontological conjunct about the
notion of truth as correspondence, and an independent
reality; and an epistemic conjunct of `optimism'---about
our mature theories `living up' to the first conjunct's
notions.

Obviously, discussions of quantum gravity (of any of the
three types of Section \ref{SubSec:NoData}) need not be
committed to such a doctrine, simply because, whatever
exactly `successful' and `mature' mean, quantum gravity
hardly supplies us with such theories! That is: even if one
endorses the first conjunct of scientific realism, the
second conjunct does not apply to quantum gravity. So there
is no commitment, whatever one's view of the first
conjunct.

But there is another way in which we might seem to be
committed to scientific realism: namely, through our
treatment of the `ingredient theories', quantum theory and
general relativity. In Section \ref{Sec:ConceptProb} {\em
et seq.\/} we will often write about the interpretation of
these theories, in an ontological (rather than
epistemological or methodological) sense; as does much
current work in the philosophy of physics. For example, we
will mention the so-called `interpretations' of quantum
theory (Copenhagen, Everettian, pilot-wave etc.); which are
in fact ontologies, or world-views, which the philosopher
of quantum theory elaborates and evaluates. Similarly as
regards general relativity; for example, we will sometimes
write about the existence of spacetime points as if they
were objects.

But it should not be inferred from our writing about these
topics in this way that we are committed to some form of
scientific realism. There is no such entailment; for two
reasons, one relating to each of scientific realism's two
conjuncts. The first, obvious reason concerns the epistemic
optimism of scientific realism. Clearly, elaborating and
evaluating ontologies suggested by scientific theories is
quite compatible with denying this optimism.

The second reason is perhaps less obvious. We maintain that
such elaboration and evaluation of ontologies involves no
commitment to a correspondence notion of truth, or
approximate truth, characteristic of realism. This can seem
surprising since for philosophers of physics, `electron'
and `spacetime point' come as trippingly off the tongue, as
`chair' and other words for Austin's `medium-sized dry
goods' come off all our tongues, in everyday life. And this
suggests that these philosophers' account of reference and
truth about such topics as electrons is as realist, as is
the account by the so-called `commonsense realist' of
reference to chairs, and of the truth of propositions about
chairs. But the suggestion is clearly false. Whatever
general arguments (for example, about ontological
relativity) can be given against realist accounts of
reference and truth in regard to chairs (and of course,
rabbits and cats---Quine's and Putnam's `medium-sized wet
goods'!) can no doubt also be applied to electrons and
spacetime points. Indeed, if there is to be a difference,
one expects them to apply with greater, not lesser, force;
not least because---at least, in the case of the
electron---of the notorious difficulty in understanding a
quantum `thing' in any simple realist way.

Here we should add that in our experience, philosophers of
physics do {\em in fact\/} tend to endorse realist accounts
of reference and truth. We suspect that the main cause of
this is the powerful psychological tendency to take there
to be real physical objects, corresponding in their
properties and relations to the mathematical objects in
mathematical models, especially when those models are very
successful. But this tendency is a cause, not a reason;
{\em i.e.\/}, it does not support the suggestion we denied
above, that elaborating a physical theory's ontology
implies commitment to realism. Whitehead had a vivid phrase
for this tendency to reification: `the fallacy of misplaced
concreteness'.

For this paper, the main example of this psychological urge
will be the tendency to reify spacetime points, which we
shall discuss in more detail in Section
\ref{SubSec:InterpretGR}. For now, we want just to make
three general points about this tendency to reification.
The third is more substantial, and so we devote the next
Subsection to it.

First, such reification is of course a common syndrome in
the praxis of physics, and indeed the rest of science;
carried over, no doubt, from an excessive zeal for realism
about say, chairs, in everyday life. Certainly, in so far
as they take a view on these matters, the great majority of
physicists tend to be straightforward realists when
referring to electrons, or even such exotic entities as
quarks.

Second, reification is not {\em just\/} a psychological
tendency, or a pedagogic crutch. It can also be
heuristically fruitful, as shown by successful physical
prediction based on the mathematics of a theory; for
example, Dirac's prediction of the positron as a `hole' in
his negative energy `sea' (though he at first identified
the holes with protons!).

\subsubsection{The Fragility of Ontology in Physics}
Setting aside our general cautiousness about scientific
realism, there is a specific reason to be wary of misplaced
concreteness in theoretical physics. We cannot develop it
fully here, but we must state it; for it applies in
particular to such putative objects as spacetime points,
which will of course be centre-stage in this paper.

The reason arises from the idea that physics aims to supply
a complete description of its subject-matter. It does not
matter how exactly this idea is made precise: for example,
what exactly `complete description' means, and whether this
aim is part of what we mean by `physics'. The rough idea of
physics aiming to be complete is enough. For it entails
that in physics, or at least theoretical physics, a change
of doctrine {\em about\/} a subject-matter is more
plausibly construed as a change of subject-matter {\em
itself\/}, than is the case in other sciences. So in
physics (at least theoretical physics), old ontologies are
more liable to be rejected in the light of new doctrine.

We can make the point with a common-sense example. Consider
some body of common-sense doctrine, say about a specific
table, or tables in general. Not only is it fallible---it
might get the colour of the specific table wrong, or it
might falsely say that all tables have four legs---it is
also bound to be incomplete, since there will be many facts
that it does not include---facts which it is the business
of the special sciences, or other disciplines, to
investigate; for example, the material science of wood, or
the history of the table(s). Similarly for a body of
doctrine, not from common-sense, but from a science or
discipline such as chemistry or history, about any
subject-matter, be it wood or the Napoleonic wars. There
are always further facts about the subject-matter, not
included in the doctrine. Indeed this is so, even if the
body of doctrine is the conjunction of all the facts about
the subject-matter expressible in the taxonomy (vocabulary)
of the discipline concerned. Even for such a giant
conjunction, no enthusiast of such a science or discipline
is mad enough, or imperialist enough, to believe that it
gives a complete description of its subject matter. There
are always other conjuncts to be had, typically from other
disciplines.

Not so, we submit, for physics, or at least theoretical
physics: whether it be madness, imperialism, or part of
what we mean by `physics', physics {\em does\/} aspire to
give just this sort of complete description of its
subject-matter. And this implies that when `other conjuncts
arrive'---{\em i.e.\/}, new facts come to light additional
to those given by the most complete available physical
description---it is more reasonable to construe the new
facts as involving a change of subject-matter, rather than
as an additional piece of doctrine about the old
subject-matter.

Note that we do {\em not\/} say that the first construal is
always more reasonable than the second---by no means! Only
that it is usually more reasonable than in other sciences,
simply because of physics' aspiration to completeness. To
take an obvious example: the very fact that a
quantum-theoretic description of the electron before the
discovery of spin aspired to be complete, makes it more
reasonable to construe the discovery of the magnetic moment
of the electron as a change of subject-matter---the
replacement of the `old' ontology comprising the spinless
electron, by one with a spinning electron---rather than as
just additional doctrine about the old subject-matter, the
`old electron'. To sum up this point: the fact that physics
aspires to give a complete description of its
subject-matter gives a specific reason to be wary of
reifying the objects postulated by physical
theories.\footnote{As mentioned, lack of space prevents a
full defence of this point. Suffice it here to add two
comments. (1) The point does not assume that new doctrine
which does {\em not\/} change the subject-matter must be
cumulative, {\em i.e.\/}, must not contradict old doctrine.
Suppose doctrine can be withdrawn or adjusted, without the
subject-matter changing: the point still holds good. (2)
The point is independent of physicalism. For it turns on
physics aspiring to give a complete description of its
subject-matter. But this implies nothing about whether the
subject-matter of physics exhausts all (empirical)
subject-matters, {\em i.e.\/}, about whether physicalism is
true.}

\subsubsection{The Question of Transcendental Idealism}
\label{SubSubSec:TranscendIdealism} Any discussion about
realism---even one mainly concerned, as we are, with
scientific realism---raises the issue of transcendental
idealism: that is, as we understand it, the issue whether
there is, or must be thought to be, something beyond the
`appearances'---or in more modern terms, the `ontology of
science'. This issue comes to mind all the more readily in
a discussion of realism concerning space and time, since it
was in connection with these categories that Kant forged
his transcendental idealism.

Of course we have no space here to address this enormous
issue.\footnote{After all, even if one confines oneself to the topic
of space and time, and to authors who seem to deserve the
name `transcendental idealists', there are several
positions to evaluate; for example, Kant's view of space
and time as ideal---the {\em a priori\/} conditions for all
experience---and Kuhn's view of the essential reality of
space.} Instead, we confine ourselves to two short remarks.
First: Like most authors in the philosophy of physics, and
almost all theoretical physicists\footnote{In fact, the
theoretical physicist (CJI) in this collaboration does {\em
not\/} hold this view.}, we will write for convenience and
brevity `as if' transcendental realism is true: more
precisely, as if there is nothing beyond `the ontology of
science'. But we stress that `the ontology of science' need
not be understood in terms of the claims of our best, or a
mature, scientific theory. For example, the phrase can be
understood in terms of a Peircean limit of enquiry; with no
implication that our best, or mature, scientific theories
approximate this limit. So whether or not transcendental
realism is true, it is in any case consistent with
rejecting scientific realism.

Second, a remark specific to quantum gravity that relates
to the minuscule size of the Planck length, emphasised at
the start of Section \ref{SubSec:NoData}. Namely, it is
{\em so\/} minuscule as to suggest that those aspects of
reality that require a theory of quantum gravity for their
description do not deserve such names as `appearance',
`phenomenon', or `empirical'. Agreed, there is no hint in
the writings of Kant or other Kantians that one should
restrict the word `appearance' to what is {\em
practically\/} accessible. And one naturally thinks that an
`item' (event, state of affairs, call it what you will)
that is localized in spacetime, or that somehow has aspects
localized in spacetime, is {\em ipso facto\/} an
appearance, part of empirical reality---be it, or its
aspects, ever so small. But we would like to suggest that
one should resist this, and to consider taking the
inaccessibility of these scales of length, energy etc.\ to
be {\em so\/} extreme as to be truly `in principle'. To put
the point in terms of `empirical': the suggestion is that
these items, or their localized aspects, are not empirical,
though one might still call them `physical', as well as, of
course, `real' and `actual'---in particular, these items or
aspects would be represented in our theory of quantum
gravity.\footnote{The suggestion also allows what we hoped
for in Section \ref{SubSec:NoData}: namely, quantum gravity
effects at much more accessible length scales; it only
contends that aspects we cannot thus probe are not
empirical.}

If this is right, one could perhaps reconcile various
Kantian claims that space and time must have some
features---for example, being continua---as an {\em a
priori\/} matter with the claims of those quantum gravity
programmes that deny space and time those features. The
apparent contradiction would be an artefact of an ambiguity
in `space and time': the quantum gravity programmes would
{\em not\/} be about space and time in the Kantian sense.
Finally, we should emphasize that in envisaging such a
reconciliation, we are not trying to defend specific
Kantian claims, such as its being an {\em a priori\/}
matter that space and time are continua; or that geometry
is Euclidean. Indeed, we join most physicists in being
sceptical of such specific claims, not least because the
history of physics gives remarkable examples of the
creative, albeit fallible, forging of new concepts. But we
{\em are\/} sympathetic to the broader Kantian idea that
human understanding of reality must, as an {\em a priori\/}
matter, involve certain notions of space and time.

\section{Conceptual Problems of Quantum Theory and General
Relativity} \label{Sec:ConceptProb} As discussed in Section
\ref{SubSec:Prologue}, the over-arching question of this
paper is: What part (if any) of the ingredient theories' common
treatment of spacetime---{\em i.e.}, as a differentiable manifold
with a Lorentzian metric---needs to be given up in quantum
gravity? It is already clear (sad to say!) that there is no
agreement about the answer to this question. As we shall see in
more detail in Sections 3 et seq., there is a wide variety of
different quantum gravity programmes, giving different answers.
And more confusingly, these different answers do not always
represent simple disagreements between the programmes: sometimes
two programmes are aiming to do such very different things, that
their different answers need not contradict each other. Hence
this paper's project of undertaking a survey.

But as we shall also see, this variety of programmes, and
of aims, is due in part to the fact that significant
conceptual problems about the ingredient theories are still
unsolved: both problems about the nature of quantum
reality, and problems about the nature of space and
time---in part, traditional philosophical problems, though
of course modified in the light of general relativity and
quantum theory. So it will help to set the stage for our
survey, to devote this Section to describing such problems.
Of course, we cannot give a thorough discussion, or even an
agreed complete {\em list\/}, of these
problems.\footnote{For more discussion, see Parts 3 and 5
of this volume.} We confine ourselves to briefly discussing
some issues that are specifically related to quantum
gravity.

In this discussion, we will place the emphasise on problems
of quantum theory, for two reasons; only the second of
which concerns quantum gravity. First, we agree with the
`folklore' in the philosophy of physics that quantum theory
faces more, and worse, conceptual problems than does
general relativity. In a nutshell: general relativity
is a classical field theory (of gravitation); and broadly
speaking, such theories are not mysterious, and their
interpretation is not controversial. On the other hand,
quantum theory {\em is\/} mysterious, and its
interpretation {\em is\/} controversial. This is attested
not only by the struggles of its founding fathers; but also
with the ongoing struggles with issues such as the
non-Boolean structure of the set of properties of a
physical system, the lack of values for quantities
associated with superpositions, the phenomenon of quantum
entanglement---and of course, the `confluence' of these
three issues in the `measurement problem'. Indeed, thanks
to these struggles, the issues are not nearly so
intractable as they were 70, or even 40, years ago. Though
mystery remains, there are nowadays several flourishing
schools of thought about how to interpret quantum theory:
we will mention some of them in Section
\ref{SubSec:InterpretQT} below.

Second, as we shall see in Section \ref{Sec:ResearchProg}:
despite general relativity's `merits' of interpretative
clarity over quantum theory, the main quantum gravity
programmes tend to put much more pressure on the framework
of standard general relativity, and thus on spacetime
concepts, than they do on quantum theory. Like most other
research programmes using quantum theory, they simply use
the standard quantum theoretic formalism, and do not
address its conceptual problems. It is this disparity that
motivates this paper's choice of spacetime as the main
topic of its survey. But arguably, this acceptance of
standard quantum theory is a mistake, for two reasons.
First: in general, it would seem wise for a research
programme that aims to combine, or somehow reconcile, two
theories, to rely more heavily on the clearer ingredient
theory, than on the mistier one! Second: as we shall see in
Section \ref{SubSec:InterpretQT}, it turns out that in
various ways, the search for a quantum theory of gravity
raises the conceptual problems of quantum theory in a
particularly acute form---and even puts some pressure on
its mathematical formalism. In any case---whether or not
this acceptance of quantum theory is a mistake---in this
Section we will briefly `redress the balance'. That is to
say, we will emphasise the pressure that quantum gravity
puts on quantum theory.\footnote{For more such pressure,
see Part 3 of this volume.}

Accordingly, our plan will be to discuss first (in Section
\ref{SubSec:InterpretQT}) the conceptual problems of
quantum theory, especially in relation to quantum gravity;
and then the conceptual problems of general relativity (in
a shorter Section \ref{SubSec:InterpretGR}).

\subsection{Interpreting Quantum Theory}
\label{SubSec:InterpretQT} In this Subsection, our strategy
will be to distinguish four main approaches to interpreting
quantum theory, in order of increasing radicalism; and to
show how each relates to topics, or even specific
approaches, in quantum gravity. The first two approaches
(discussed in Sections \ref{SubSubSec:Instrumentalism} and
\ref{SubSubSec:Literalism}) are both conservative about the
quantum formalism---they introduce no new equations. But
they differ as to whether they are `cautious' or
`enthusiastic' about the interpretative peculiarities of
quantum theory. The third and fourth approaches (discussed
in Sections \ref{SubSubSec:ExtraValues} and
\ref{SubSubSec:NewDynamics}) do introduce new equations;
but in various (different) ways remain close enough to
standard quantum theory to be called `interpretations' of
it.

We stress that although our catalogue of four approaches is
by no means maverick, we make no claim that it is the best,
let alone the only, way to classify the various, and
complexly inter-related, interpretations of quantum theory.
But of course we believe that with other such
classifications, we could make much the same points about
the connections between interpreting quantum theory and
quantum gravity.

On the other hand, we cannot consider the details of
individual interpretations within each approach. And in
view of the paper's overall project, we shall emphasise how
the interpretations we do mention relate to space and
time---at the expense of other aspects of the
interpretation. For example, we will not mention even such
basic aspects as whether the interpretation is
deterministic---except in so far as such aspects relate to
space and time.

\subsubsection{Instrumentalism}\label{SubSubSec:Instrumentalism}
We dub our first approach to interpreting quantum theory,
`instrumentalism'. We intend it as a broad church. It
includes views that apply to quantum theory some general
instrumentalism about all scientific theories; and views
that advocate instrumentalism only about quantum theory,
based on special considerations about that subject. We will
not comment on the first group, since we see no special
connections with quantum gravity. Or more precisely, we see
no connections other than those which we already adumbrated
from another perspective, that of realism, in Section
\ref{Sec:Intro}, especially Section \ref{SubSec:Realism?}.

On the other hand, some views in the second group do have
connections with quantum gravity, albeit `negative' ones.
Thus consider the Copenhagen interpretation of quantum
theory: understood, not just as the minimal statistical
interpretation of the quantum formalism in terms of
frequencies of measurement results, but as insisting on a
classical realm external to the quantum system, with a firm
`cut' between them, and with no quantum description of the
former. In so far as this classical realm is normally taken
to include classical space and time\footnote{Kantian themes
about the {\em a priori\/} nature of space and time arise
here, just as in Section \ref{SubSec:Realism?}; for
Kantian aspects of Bohr, see \cite{Kai92}.}, this suggests
that, in talking about `quantum gravity', we are making a
category error by trying to apply quantum theory to
something that forms part of the classical background of
that theory: ``what God has put assunder, let no man bring
together''. We shall say more later about the view that a
quantum theory of gravity should, or can, be avoided
(Section \ref{SubSubSec:QGAvoided?}). But for the most part
we will accept that serious attempts should be made to
construct a `quantum theory of space and time' (or, at
least, of certain aspects of space and time); with the
understanding that, in doing so, it may be necessary to
radically change the interpretation---and, perhaps, the
mathematical formalism---of quantum theory itself.

In endeavouring to interpret quantum theory, regardless of
quantum gravity, this second group of views is notoriously
problematic. It is not just a matter of it being difficult
to understand, or to defend, Bohr's own views, or views
similar to his. There are quite general problems, as
follows. Any view that counts as `instrumentalism
specifically about quantum theory' ({\em i.e.}, any view in
this group) must presumably do either or both of the
following:
\begin{enumerate}
\item[(i)] deny that the quantum state describes
individual systems, at least between measurements; or in
some similar way, it must be very cautious about the
quantum description of such systems;

\item[(ii)] postulate a `non-quantum' realm, whose
description can be taken literally ({\em i.e.\/}, not
instrumentalistically, as in (i)); usually this realm is
postulated to be `the classical realm', understood as
macroscopic, and/or the domain of `measurement results',
and/or described by classical physics.
\end{enumerate}
But recent successful applications of quantum theory to
individual microphysical systems (such as atoms in a trap),
and to macroscopic systems (such as superconducting squids)
have made both (i) and (ii) problematic. This suggests in
particular that we should seek an interpretation in which
no fundamental role is ascribed to `measurement',
understood as an operation external to the domain of the
formalism; see the next Subsection.

\subsubsection{Literalism}\label{SubSubSec:Literalism}
Like instrumentalism, we intend this approach to be a broad
church. The idea is to make the interpretation of quantum
theory as `close' as possible to the quantum formalism.
(Hence the name `literalism'; `realism' would also be a
good name, were it not for its applying equally well to our
third and fourth approaches.) In particular, one rejects
the use of a primitive notion of measurement, and
associated ideas such as a special `classical realm', or
`external observer' that is denied a quantum-theoretic
description. Rather, one `cuts the interpretation to suit
the cloth of the formalism'; revising, if necessary,
traditional philosophical opinions, in order to do so.
Hence our remark at the start of Section
\ref{SubSec:InterpretQT} that this approach is
`enthusiastic' about the interpretative peculiarities of
quantum theory, while instrumentalism is `cautious'.

As we see it, there are two main types of literalist view:
Everettian views, and those based on quantum logic. Of
these types, the first has been much discussed in
connection with quantum gravity (especially quantum
cosmology); but the second, hardly at all in this
connection. Accordingly, we shall only treat the
first.\footnote{About the second, suffice it to say that
such views propose to revise the logic of discourse about
quantum systems; (so we do not intend the type to include
purely technical investigations of non-Boolean structures).
But it is unclear how such proposals solve the
interpretative problems of quantum theory, such as the
measurement problem or non-locality; and indeed, these
proposals now seem to have few advocates---at least
compared with 25 years ago. In any case, their advocates
hardly connect them with quantum gravity.}

As usually presented, the main aim of an Everettian view
(or, as it is sometimes called these days, a
`post-Everrett' view) is to solve the `measurement
problem': {\em i.e.\/}, the threat that at the end of a
measurement, macroscopic objects (such as an instrument
pointer) will have no definite values for familiar
quantities like position---contrary to our experience. More
specifically, the aim is to solve this problem without
invoking a collapse of the state vector, or an external
observer. This involves (i) resolving the state vector of a
closed system as a superposition of eigenstates of a
`preferred quantity'; (ii) interpreting each of the
components as representing definite positions for pointers
and other macroscopic objects; and then (iii) arguing that,
although there is no collapse, you will only `see' one
component in the superposition.

This summary description leaves open some crucial
questions. For example:
\begin{enumerate}
\item[(a)] How is the preferred quantity to be chosen?
Should it be in terms of familiar quantities such as
position of macroscopic objects, so that each summand
secures a definite macroscopic realm (`many worlds'); or
should it involve arcane quantities concerning brains,
whose eigenstates correspond to experiences of a definite
macroscopic realm (`many minds')?

\item[(b)] Should one say that for each component there is a physically real
`branch', not just the possibility of one?

\item[(c)] How should one justify the claim that you will not `see the other
components': by some process of `splitting of the
branches', or by appeal to decoherence making the
interference terms that are characteristic of the presence
of other components, negligibly small?
\end{enumerate}
We do not need to discuss these issues here, which have
been much debated in the philosophy of quantum
theory.\footnote{We do so, albeit briefly, in \cite{BI99a};
see also \cite{Butt95}, and \cite{Butt96}.} For our
purposes, it suffices to note the four main connections of
Everettian views with quantum gravity, specifically quantum
cosmology.

The first connection has been evident from the earliest
discussions of Everettian views. Namely, whatever the exact
aims of a theory of quantum cosmology, in so far as it
posits a `quantum state of the universe', the Everettian
promise to make sense of the quantum state of a closed
system makes this interpretation particularly attractive.

The second connection concerns the more extreme Everettian view
in which the universe is deemed literally to `split'. In so far
as this might involve some transformation of the topology of
space, one naturally imagines implementing this with the aid of
ideas from quantum gravity.

The third connection relates to decoherence, mentioned in
question (c) above. Much recent work has shown decoherence
to be a very efficient and ubiquitous mechanism for making
interference terms small (and so for securing an apparent
reduction of a quantum system's state vector); essentially
by having the correlational information that these terms
represent `leak out' to the system's environment. Though
this work in no way relies on Everettian views, Everettians
can, and do, appeal to it in answering question (c).
Furthermore, the work has been adapted to the discussion
within quantum cosmology of how we `see' a single classical
space or spacetimes, despite the fact that in quantum
cosmological models the quantum state of the universe
assigns non-zero amplitude to many such spaces. The idea is
that decoherence destroys the interference terms, `hiding
all but one'. (Typically, inhomogeneous modes of the
gravitational field act as the environment of the
homogeneous modes, which form the system.) For more
discussion, see \cite{Rid99b}.

The fourth connection relates to time. Any Everettian view
must specify not only probabilities for values of its
preferred quantity at each time; it must also specify joint
(`conjunctive') probabilities for values at sequences of
times; {\em i.e.\/}, a rule for the temporal evolution of
these values. Traditionally, Everettians tended not to give
such a rule; but recently they have done so, often in the
context of the `consistent-histories' approach to quantum
theory.

There is a specific reason for quantum cosmologists to
focus on the consistent-histories formalism, apart from the
general need to specify a rule for the evolution of values.
As we shall see in more detail in Section
\ref{SubSec:ProblemTime}, quantum gravity, and thereby
quantum cosmology, is beset by `the problem of time'. One
response to this severe problem is to seek a new type of
quantum theory in which time does not play the central role
that it does in the standard approach. And precisely
because the consistent-histories approach concerns many
times, it suggests various ways in which the formalism of
quantum theory can be generalised to be less dependent on
the classical concept of time \cite{Har95,IL94}.

This last point gives an example of an important, more
general idea, which goes beyond the discussion of
Everettian views. Namely, it is an example of how issues in
quantum gravity can put pressure on the actual formalism of
quantum theory---not just on some traditional
interpretative views of it, such as the Copenhagen
interpretation.

\subsubsection{Extra Values}\label{SubSubSec:ExtraValues}
 Again, we intend this approach to be a broad church.
Like the Everettian views discussed above, it aims to
interpret quantum theory---in particular, to solve the
measurement problem---without invoking a collapse of the
state vector. And it aims to do this by postulating values
for some `preferred quantity' or quantities, in addition to
those given by the orthodox `eigenvalue-eigenstate
link'\footnote{This asserts that a system has a real number
$r$ as its value for a quantity $Q$ if and only if the
quantum state is an eigenstate of $Q$ with eigenvalue
$r$.}; together with a rule for the evolution of such
values.

But there are two differences from the Everettian views.
First, `Extra Values' makes no suggestion that there is a
physically real `branch' for every component in the
resolution of the state vector in terms of the preferred
quantity. (So there is no suggestion that `branches
splitting' prevents the detection of interference terms.)
Second, `Extra Values' aspires to be more precise from the
outset about which quantity is preferred, and the dynamics
of its values.\footnote{Agreed, this second difference is a
matter of degree. Furthermore, Everettians' imprecision
about the preferred quantity, and the dynamics of its
values, is partly just an accident, due to the facts that:
(i) their view first developed within traditional quantum
measurement theory, which invokes imprecise notions like
`apparatus' and `pointer-position'; and (ii) they are
willing to secure only the appearance of a definite
macroscopic realm, not a truly definite one, and are
therefore able to leave future psychophysics to specify the
preferred quantities. But this matter of degree is no
problem for us: our taxonomy of four approaches is not
intended to be rigid.}

The best-known examples of this approach are the
deBroglie-Bohm `pilot-wave' or `causal' interpretation of
quantum theory \cite{Val96}; and the various kinds of modal
interpretation \cite{Bub97}. Thus the pilot-wave
interpretation of quantum mechanics postulates a definite
value for the position of each point-particle, evolving
according to a deterministic guidance equation. The
corresponding interpretation of quantum field theory
postulates a definite field configuration, and again a
deterministic guidance equation. On the other hand, modal
interpretations postulate that which quantity is
`preferred' depends on the state; and they consider various
stochastic dynamics for values.

Within this approach, only the pilot-wave interpretation
has been discussed in connection with quantum gravity. The
main idea is to `make a virtue of necessity', as
follows.\footnote{For more discussion, see the Chapter in
this volume by Goldstein, and \cite{Val96}.} On the one hand, the guidance
equations (at least as developed so far) require an
absolute time structure, with respect to which the
positions or field configurations evolve. (So for familiar
quantum theories on Minkowski spacetime, the relativity of
simultaneity, and the Lorentz-invariance of the theory, is
lost at the fundamental level---but recovered at a
phenomenological level.) On the other hand: in quantum
gravity, one response to the problem of time is to `blame'
it on general relativity's allowing arbitrary foliations of
spacetime; and then to postulate a preferred foliation of
spacetime with respect to which quantum theory should be
written. Most general relativists feel this response is too
radical to countenance: they regard foliation-independence
as an undeniable insight of relativity. But an advocate of
the pilot-wave interpretation will reply that the virtues
of that interpretation show that sacrificing fundamental
Lorentz-invariance is a price worth paying in the context
of flat spacetime; so why not also `make a virtue of
necessity' in the context of curved spacetime, {\em
i.e.\/}, general relativity?

Indeed, this suggestion has been developed in connection
with one main approach to quantum gravity; namely, the
quantum geometrodynamics version of the canonical quantum
gravity programme. We will discuss this in more detail in
Section \ref{SubSec:CanWQG}. For the moment, we just note
that the main idea of the pilot-wave interpretation of
quantum geometrodynamics is to proceed by analogy with the
interpretation of quantum field theories such as
electrodynamics on flat spacetime. Specifically, a
wave-function defined on 3-geometries (belonging to the
3-dimensional slices of a preferred foliation) evolves in
time, and deterministically guides the evolution of a
definite 3-geometry.

To sum up: `Extra Values' preserves the usual unitary
dynamics (the Schr\"{o}dinger equation) of quantum theory,
but adds equations describing the temporal evolution of its
extra values. And the best developed version of `Extra
Values'---the pilot-wave interpretation---has been applied
only to the quantum gravity programme based on quantum
geometrodynamics.

\subsubsection{New Dynamics}\label{SubSubSec:NewDynamics}
This approach is more radical than Extra Values. Instead of
adding to the usual unitary dynamics of quantum theory, it
replaces that dynamics; the motivation being to solve the
measurement problem by dynamically suppressing the
threatened superpositions of macroscopically
distinguishable states. In the last fifteen years, there
has been considerable development of this approach,
especially in the wake of the `spontaneous localization'
theories of Ghirardi, Rimini and Weber \cite{GRW86}, and
Pearle \cite{Pea89}.

This approach has natural links with quantum gravity.
Indeed, from the point of view of physical theory itself,
rather than its interpretation, it is a closer connection
than those reviewed in the previous Subsections. For it is
natural to suggest that the proposed deviation from the
usual dynamics be induced by gravity (rather than being
truly `spontaneous'). This is natural for at least two
reasons: (i) gravity is the only {\em universal\/} force we
know, and hence the only force that can be guaranteed to be
present in all physical interactions; and (ii)
gravitational effects grow with the size of the objects
concerned---and it is in the context of macroscopic objects
that superpositions are particularly problematic.

We emphasise that this idea---that gravity is involved in
the reduction of the state vector---is different from, and
more radical than, the idea in Subsection
\ref{SubSubSec:Literalism} that some modes of the
gravitational field might act as the environment of a
system for a decoherence process that yields an apparent
state-vector reduction. Here, there is no invocation of an
environment; {\em i.e.}, there is reduction for a strictly
isolated system.

This idea has been pursued in various ways for several
decades. In particular, adapting the idea to {\em
quantum\/} gravity: since general relativity treats gravity
as spacetime curvature, the most straightforward
implementation of the idea will require that a quantum
superposition of two spacetime geometries, corresponding to
two macroscopically different distributions of mass-energy,
should be suppressed after a very short time. Penrose has
been particularly active in advocating this idea. More
specific implementations of the idea involve variants of
the spontaneous localization theories; for example, Pearle
and Squires \cite{PS95}, which also contains a good
bibliography.\footnote{For further discussion and
references, see the Chapter by Christian in this volume.}

\subsection{Interpreting General Relativity}
\label{SubSec:InterpretGR} We turn now to consider the
conceptual problems of general relativity, especially those
related to quantum gravity. However, our discussion will be
briefer than that of Section \ref{SubSec:InterpretQT}, for
the two reasons given at the start of the Section. First,
general relativity is essentially a classical field theory,
and its interpretation is less mysterious and controversial
than that of quantum theory. Second, subsequent Sections
will give ample discussion of the pressure that quantum
gravity puts on general relativity.

Specifically, we shall confine ourselves to brief remarks
about one aspect of the grand debate between `absolute'
{\em versus\/} `relational' conceptions of space and time:
namely, the question in what sense, if any, spacetime
points are objects.\footnote{We should stress that here we
take the word `object' in the `post-Fregean' sense of
anything that could be the referent of a singular term. In
particular, it does not necessarily mean a `thing out
there' as a physicist might construe the phrase.}(We shall
discuss spacetime points, but most of the discussion could
be straightforwardly rephrased as about whether regions in
spacetime are objects; or, indeed---in a canonical
approach---points or regions in $3$-space, or in time.) For
this question bears directly on the discussion in
subsequent Sections of the treatment of spacetime in
quantum gravity.\footnote{Agreed, there are many other
conceptual aspects of general relativity that bear on
quantum gravity. We mentioned the philosophy of geometry at
the end of Section \ref{SubSec:NoData}. Another obvious
example is the global structure of time, which bears on
quantum cosmology. Here one faces such issues as: In what
sense could a quantum event `precede' the big-bang? This is
related to the `problem of time' in quantum gravity; which
will be discussed later (Section \ref{SubSec:ProblemTime});
see also Part 2 of this volume, and Section 5 of our essay
\cite{BI99a}). For a philosophical discussion of several
other conceptual aspects of general relativity bearing on
quantum gravity, see \cite{Ear95}.}

The debate between `absolute' and `relational' conceptions
of space and time has many strands. Nowadays, philosophers
separate them, at least in part, by distinguishing various
senses. For example, does a spatiotemporal structure being
`absolute' mean that it is `non-dynamical', {\em i.e.},
unaffected by material events; or that it is not determined
by (supervenient on) the spatiotemporal relations of
material bodies?\footnote{In any case, the rise of field
theory undermines the contrast one naively learns in
everyday life, between empty space and material bodies.}
And what spatiotemporal structure does the `absolutist'
take to be absolute: space (as by Newton), or the
4-dimensional metric of spacetime, or the connection? Once
these senses are distinguished, it becomes clear that
general relativity supports `relationism' in the senses
that (i) its 4-dimensional metric and connection are
dynamical; and (ii) in its generic models, no space, {\em
i.e.\/}, no foliation of spacetime, is preferred (whether
dynamically or non-dynamically). On the other hand, it
supports absolutism in the sense that the {\em presence\/}
in the theory of the metric and connection is not
determined by the spatiotemporal relations of material
bodies.

But the consensus on these issues about relatively
technical senses of `absolute' leaves outstanding the
question whether we should interpret general relativity as
commited to the existence of spacetime points (or regions)
as physical objects. We are wary of the `Yes' answer to this
question (which became popular in the 1960s, with the rise
of scientific realism). But this is not just because we are
wary of scientific realism; and in particular, of reifying
the objects and structures postulated by physical theories
(as discussed in Section \ref{SubSec:Realism?}). We also
have two more specific reasons. The first contains a more
general moral about reification; the second is specific to
spacetime points.

(1): To explain the first reason, we should begin by admitting
that it is especially tempting to take spacetime points as
the fundamental physical objects of both our `ingredient
theories'---general relativity and quantum theory---and not
just as points in mathematical models. There are two
specific factors prompting this reification.

Roughly speaking, the first factor is this. As usually
formulated, the theories agree with one another in postulating
such points, endowed with the (highly sophisticated) structure of
a differentiable manifold. But to be precise, one needs to
respect the distinction between a (putative) physical spacetime
point, and an (undeniably postulated!) point in a mathematical
model of spacetime. So one should express this first factor by
saying that, as usually formulated, the theories agree in
postulating the latter points, {\em i.e.\/}, those in the
mathematical models. Even the most ardent realist must allow this
distinction in principle, if she is to avoid begging the
question; though she may well go on to suggest that as a realist,
she can take (and perhaps prefers to take) the physical points in
which she believes, as elements of the mathematical model---say
as the bottom-level elements in a set-theoretic definition of a
manifold equipped with a Lorentzian metric and some matter
fields.

The second factor is that, as usually formulated, the
theories postulate the points initially, {\em i.e.\/}, at
the beginning of their formalism; the rest of physical
reality being represented by mathematical structures
(vector, tensor and operator fields etc.) defined over the
points.\footnote{This of course reflects the rise, from the
mid-nineteenth century onwards, of the field-theoretic
conception of matter, whether classical or quantum. From
another perspective, it reflects the dominant position of
set theory in the foundations of mathematics.} Here, as in
the first factor, to be precise---and to avoid begging the
question---one must take these postulated points to be
those in the mathematical models, not the putative physical
points. The other structures representing fields etc.\ then
become properties and relations among these postulated
points; or, more generally, higher-order properties and
relations; or in a formal formulation of the theory,
set-theoretic surrogates for such properties and relations.
And again, to avoid begging the question one must
understand `represent' as not committing one to the
represented fields {\em really\/} being properties and
relations. For that would commit one to there being objects
which instantiated them; and these would no doubt be
spacetime points and $n$-tuples of them.

However, notwithstanding these cautionary remarks, most
people who bother to think about such matters succumb to
Whitehead's fallacy of misplaced concreteness, by positing
a one-to-one correspondence between what is undeniably real
in the Platonic realm of mathematical form, and what is,
more problematically, `real' in the world of physical
`stuff'.

But tempting though this reification is, it is very
questionable: not least because it overlooks the fact that
these theories can be formulated in {\em other\/} (usually
less well-known) ways, so as to postulate initial
structures that, from the usual viewpoint, are complex
structures defined {\em on\/} the points.

More precisely, the theories can be formulated so as to
postulate initially not (i) mathematical objects that
represent spacetime points (again, understanding
`represent' as not committing one to spacetime points being
genuine physical objects); but rather (ii) mathematical
objects that represent (again understood non-commitally)
fields, and similar items---items that in the usual
formulations are represented by complex mathematical
structures (formally, set-theoretic constructions) defined
over the initially-postulated representatives of spacetime
points.

To give the flavour of such formulations, here is a
standard example from the simpler setting of topological
spaces, rather than differentiable manifolds. Consider a
compact Hausdorff space $X$ and let $\cal A$ denote the
ring of real-valued functions on $X$. Then it is a famous
theorem in topology that both the set $X$ ({\em i.e.}, its
points) and the topology of $X$ can be uniquely
reconstructed from just the {\em algebraic\/} structure of
$\cal A$. Specifically, the closed subsets of $X$ are in
one-to-one correspondence with the (closed) ideals in the
commutative ring (actually, $C^*$-algebra) $\cal A$; and
the points of $X$ correspond to maximal ideals in $\cal A$.

The implication of this result is that, from a mathematical
perspective, a theory based on such a topological
space---modeling, say, physical space---can be formulated in such
a way that the fundamental mathematical entity is not the set $X$
of spatial points---on which {\em fields\/} are then
defined---but rather a commutative ring, on which spatial {\em
points\/} are then defined: viz.\ as maximal ideals. Put in
graphic terms, rather than writing $\phi(x)$, one writes
$x(\phi)$!

We emphasise that this idea is by no means
esoteric from the perspective of theoretical physics. For
example, the subject of `non-commutative' geometry starts
from precisely this situation and then posits a
non-commutative extension of $\cal A$. In this case the
algebra remains, but the points go in the sense that the
algebra can no longer be written as an algebra of functions
on anything.

(2): But there is also another reason for wariness about the
existence of spacetime points as physical objects; a reason
relating to symmetry transformations.  The idea goes back to
Leibniz; but in modern terms, it is as follows: given that a
model of a theory represents a physical possibility, the model
obtained by applying a global symmetry transformation to it
describes {\em the same\/} physical possibility. In the context
of spacetime theories, this idea means that taking points to be
physical objects involves a distinction without a difference. So
in particular, the existence of translation invariance in
Newtonian or Minkowski spacetime shows that points should not be
taken as physical objects. We think that most physicists would
concur with this idea.

In the context of general relativity, such considerations
become perhaps yet more convincing, in view of Einstein's
`hole argument'. We cannot enter into details about this
argument; which has interesting historical and physical, as
well as philosophical, aspects. Suffice it to say that: (i)
the argument applies not just to general relativity, but to
any generally-covariant theory postulating a spacetime
manifold; and (ii) according to the argument, general
covariance (that is: the diffeomorphism-invariance of the
theory), together with spacetime points being physical
objects, implies a radical indeterminism: and such
indeterminism is unacceptable---so that we should conclude
that points are not physical objects. That is, the points
occurring in the base-sets of differentiable manifolds with
which general relativity models spacetime should not be
reified as physically real.\footnote{This view of the
argument seems to have been Einstein's own view, from the
time of his discovery of general relativity in 1915
onwards. It was resuscitated for philosophers in
\cite{EarNor87}. For discussion and references (including
replies on behalf of points being physically real) see
\cite{Ear89}.} We shall take up this theme again in more
detail in Section \ref{SubSec:RoleDiffeos}.

\section{Introducing Quantum Gravity}\label{Sec:WhatisQG} We turn now to our main project:
surveying how quantum gravity suggests fundamental
limitations in the familiar treatment of space and time
that is common to the `ingredient theories'---quantum
theory and general relativity. In this Section, we first
give some details about the variety of approaches to
quantum gravity (Section \ref{SubSec:ApproachesQG}). Then
we give a more detailed discussion of two conceptual
aspects relating specifically to spacetime: viz.\ the role
of diffeomorphisms (Section \ref{SubSec:RoleDiffeos}) and
the problem of time (Section \ref{SubSec:ProblemTime}).
This will set the stage for the discussion in Section
\ref{Sec:ResearchProg} of the treatment of spacetime in
three of the main research programmes in quantum gravity.

\subsection{Approaches to Quantum Gravity}
\label{SubSec:ApproachesQG} In this Subsection, we begin to
give a more detailed picture of quantum gravity research.
We will first survey some motivations for studying quantum
gravity (Section \ref{SubSubSec:MotivationsQG}). Then we
will consider, but reject, the view that quantum gravity
can be avoided (Section \ref{SubSubSec:QGAvoided?}). Then
we will describe four broad approaches to quantum gravity
(Section \ref{SubSubSec:FourApproachesQG}).

\subsubsection{Motivations for Studying Quantum Gravity}
\label{SubSubSec:MotivationsQG} In surveying quantum
gravity, it is useful to begin with the various motivations
for studying the subject. For as we have seen, quantum
gravity does not have a well-established body of `facts'
against which theories can be tested in the traditional
way. In consequence, although some people's motivations
refer to potential observations or
experiments---particularly in the area of cosmology---most
motivations are of a more internal nature: namely, the
search for mathematical consistency, or the implementation
of various quasi-philosophical views on the nature of space
and time. And, since these different motivations have had a
strong influence on researchers' technical approaches to
the subject, it is important to appreciate them in order to
understand what people have done in the past, and to be
able to judge if they succeeded in their endeavours: since
to be adjudged `successful' a theory must presumably either
point beyond itself to new or existing `facts' in the
world, or else achieve some of its own internal goals.

    It is useful pedagogically to classify
motivations for studying quantum gravity according to
whether they pertain to the perspective of elementary
particle physics and quantum field theory, or to the
perspective of general relativity. As we shall see, this
divide substantially affects one's approach to the subject,
in terms of both the goals of research and the techniques
employed.

\paragraph{Motivations from the perspective of elementary
particle physics and quantum field theory}
\begin{enumerate}
\item Matter is built from elementary particles that {\em are\/}
described in quantum theoretical terms, and that certainly
interact with each other gravitationally. Hence it is
necessary to say {\em something\/} about the interface
between quantum theory and general relativity, even if it
is only to claim that, `for all practical purposes', the
subject can be ignored; (see Section
\ref{SubSubSec:QGAvoided?} for more discussion).

\item Relativistic quantum field theory might only make proper sense
if gravity is included from the outset. In particular, the
short-distance divergences present in most such
theories---including those that are renormalisable, but not
truly finite---might be removed by a fundamental cut-off at
the Planck energy. Superstring theory (see Section
\ref{SubSec:Superstrings}) is arguably the latest claimant
to implement this idea.

\item {A related claim is that considerations about quantum
gravity will be a necessary ingredient in any
fully consistent theory of the unification of the three
{\em non}-gravitational forces of nature.\footnote{The four
`fundamental' forces recognised by present-day physicists
are the electromagnetic force; the `weak' nuclear force,
that it responsible for radioactive decay; the `strong'
nuclear force, that binds together the constituents of
nuclei; and the gravitational force.} The underlying idea
here is as follows.

The mark of unification in a field theory is the equality
of the coupling constants that determine the strengths of
the different forces. However, in the quantum version these
coupling `constants' are energy dependent (they are said to
`run' with the energy) and therefore forces that are not
unified at one energy may become so at a different one. It
turns out that the running constants of the
electromagnetic, weak and strong nuclear forces can be
shown to `meet', {\em i.e.\/}, to be equal or at least
approximately equal, at around $10^{20}$Mev. The fact that
$10^{20}$Mev is ``quite close to'' the Planck energy (viz.\
only two orders of magnitude less) then suggests that
quantum gravity may have a role to play in this unification
of forces.

This line of thought also prompts the more specific
suggestion that a successful theory of quantum gravity {\em
must\/} involve the unification of all four fundamental
forces. As we shall see in Section \ref{Sec:ResearchProg},
one of the key differences between the two most currently
active research programmes, superstring theory and
canonical quantum gravity, is that the former adopts this
suggestion---it aims to to provide a scheme that
encompasses all the forces---while the latter asserts that
a quantum theory of pure gravity {\em is \/} possible. }
\end{enumerate}

\paragraph{Motivations from the perspective of a general relativist}
\begin{enumerate}
\item Spacetime singularities arise inevitably in
general relativity if the energy and momentum of any matter
that is present satisfies certain, physically
well-motivated, positivity conditions. It has long been
hoped that the prediction of such pathological behaviour
can be removed by the correct introduction of quantum
effects.

\item A related point is that, once quantum
mechanical effects are included, black holes produce
Hawking radiation and, in the process, slowly lose their
mass. But the nature of the final state of such a system is
unclear, and much debated; providing another reason for
studying quantum gravity.

\item {Quantum gravity should play a vital role in the physics of
the very early universe, not least because, in standard
classical cosmology, the `initial' event is an example of a
spacetime singularity. Possible applications include:
\begin{enumerate}

    \item finding an explanation of why spacetime has a
macroscopic dimension of four\footnote{This does not
exclude a Kaluza-Klein type higher dimension at Planckian
scales. Indeed, superstring theory suggests strongly that
something like this does occur.};

    \item accounting for the origin of the inflationary evolution
that is believed by many cosmologists to describe the
universe as it expanded after the initial big-bang.
\end{enumerate}}

\item {Yet more ambitiously, one can hope that a theory of quantum gravity will provide a quantum cosmology, and thereby an understanding of the very origin
of the universe in the big-bang as some type of quantum
`event'.

However, special problems are posed by quantum cosmology,
for example about the interpretation of the quantum state
(see Section \ref{SubSec:InterpretQT}); so one might well
take the view that quantum gravity research should not get
distracted by debating these problems. Certainly it would
be a signal achievement to have a theory that successfully
handled quantum theory and general relativity `in the
small', even if it could not be applied to the `universe in
its totality'---a problematic concept on any view! In any
case, we shall from now on largely set aside quantum
cosmology, and its special problems.\footnote{Our
complementary essay \cite{BI99a}, discusses this in some
detail, especially as regards the Euclidean
programme---which is mentioned in Section
\ref{Sec:ResearchProg} below only as a species of canonical
quantum gravity.} }

\end{enumerate}

\subsubsection{Can Quantum Gravity be Avoided?}
\label{SubSubSec:QGAvoided?} The argument is sometimes put
forward that the Planck length
$L_P:=(G\hbar/c^3)^{\frac12}\simeq 10^{-35}\mbox{m}$ is so
small that there is no need to worry about quantum gravity
except, perhaps, in recherch\'e considerations of the
extremely early universe---{\em i.e.}, within a Planck time
($\simeq 10^{-42}\mbox{s}$) of the big-bang. However, as we
hinted in the first motivation listed in Section
\ref{SubSubSec:MotivationsQG}:
\begin{itemize}
\item Such a claim is only really meaningful if a theory exists
within whose framework genuine perturbative expansions in
$L_P/L$ can be performed, where $L$ is the length scale at
which the system is probed: one can then legitimately argue
that quantum effects are ignorable if $L_P/L\ll 1$. So we
must try to find a viable theory, even if we promptly
declare it to be irrelevant for anything other than the
physics of the very early universe.

    \item The argument concerning the size of $L_P$ neglects the
possibility of {\em non\/}-perturbative effects---an idea
that has often been associated with the claim that quantum
gravity produces an intrinsic cut-off in quantum field
theory.
\end{itemize}
\smallskip

    A very different, and less radical, view is that---although
we presumably need some sort of theory of quantum gravity
for the types of reason listed in Section
\ref{SubSubSec:MotivationsQG}---it is {\em wrong\/} to try
to construct this theory by quantising the gravitational
field, {\em i.e.\/}, by applying a quantisation algorithm
to general relativity (or to any other classical theory of
gravity). We shall develop this distinction between the
general idea of a theory of quantum gravity, and the more
specific idea of quantised version of general relativity,
immediately below (Section
\ref{SubSubSec:FourApproachesQG}). For the moment, we
mention some reasons advanced in support of this view.
\begin{itemize}
\item The metric tensor  should not be viewed as  a `fundamental' field
in physics, but rather as a phenomenological description of
gravitational effects that applies only in regimes well
away from those characterised by the Planck scale. Again,
diverse reasons can be given for this viewpoint: we cite
three. One example is superstring theory. Here, the basic
quantum entities are very different from those in classical
general relativity, which is nevertheless recovered as a
phenomenological description. Another---very
different---example is Jacobson's re-derivation of the
Einstein field equations as an equation of state
\cite{Jac95}, which---presumably---it would be no more
appropriate to `quantise' than it would the equations of
fluid dynamics.\footnote{In 1971, one of us (CJI) took part
in a public debate with John Stachel who challenged the
former on this very issue. As a keen young quantum field
theorist, CJI replied that he was delighted to quantise
everything in sight. These days, not least because of the
moderating influence of his philosopher friends, he is more
cautious!} Yet a third example is Brown's view of the
metric, even in special relativity, as phenomenological
(see his Chapter in this volume).

    \item The gravitational field is concerned with the structure of
space and time---and these are, {\em par excellence\/},
fundamentally classical in nature and mode of functioning.
As we mentioned before, this might be defended from the
viewpoint of (a version of) the Copenhagen interpretation
(Section \ref{SubSec:InterpretQT})---or even from a Kantian
perspective (Section \ref{SubSec:Realism?}).
\end{itemize}

    If it is indeed wrong to quantise the gravitational field
(for whichever of the above reasons) it becomes an urgent
question how matter---which presumably {\em is\/} subject
to the laws of quantum theory---should be incorporated in
the overall scheme. To discuss this, we shall focus on the
so-called `semi-classical quantum gravity' approach. Here,
one replaces the right-hand-side of Einstein's field
equations by a quantum expectation value, so as to couple a
classical spacetime metric $\gamma$ to quantized matter by
an equation of the form
\begin{equation}
    G_{\mu\nu}(\gamma)=
        \langle\psi|T_{\mu\nu}(g,\widehat{\phi})|\psi\rangle
                            \label{G=<T>}
\end{equation}
where $|\psi\rangle$ is some state in the Hilbert space of
the quantised matter variables $\widehat\phi$. Thus the
source for the gravitational field---{\em i.e.}, the right
hand side of Eq.\ (\ref{G=<T>})---is the expectation value
of the energy-momentum tensor $T_{\mu\nu}$ of the quantised
matter variables. In this context, we note the following:
\begin{itemize}
\item In the case of electromagnetism, the well-known analysis
by Bohr and Rosenfeld \cite{BR33} of the analogue of Eq.\
(\ref{G=<T>}) concluded that the electromagnetic field {\em
had\/} to be quantised to be consistent with the quantised
nature of the matter to which it couples. However, the
analogous argument for general relativity does not go
through, and---in spite of much discussion since then (for
example, see Page and Geilker \cite{PG81})---there is
arguably still no definitive proof that general relativity
{\em has\/} to be quantised in some way.

    \item The right hand side of Eq.\ (\ref{G=<T>}) generates a
number of technical problems. For example, the expectation
value has the familiar `ultra-violet' divergences that come
from the mathematically ill-defined short-distance
behaviour of quantum fields. Regularisation methods only
yield an unambiguous expression when the spacetime metric
$\gamma$ is time-independent\footnote{More precisely, the
spacetime metric has to be static or stationary.}---but
there is no reason why a semi-classical metric should have
this property. In addition, there have been many arguments
implying that solutions to Eq.\ (\ref{G=<T>}) are likely to
be unstable against small perturbations,
and---therefore---physically unacceptable.

    \item It is not at all clear how the state $|\psi\rangle$
is to be be chosen. In addition, if $|\psi_1\rangle$ and
$|\psi_2\rangle$ are associated with a pair of solutions
$\gamma_1$ and $\gamma_2$ to Eq.\ (\ref{G=<T>}), there is
no obvious connection between $\gamma_1$ and $\gamma_2$ and
any solution associated with a linear combination of
$|\psi_1\rangle$ and $|\psi_2\rangle$. Thus the quantum
sector of the theory has curious non-linear features, and
these generate many new problems of both a technical and a
conceptual nature.
\end{itemize}

So much by way of reviewing the reasons one might give for
avoiding quantum gravity. We make no claim that our
`replies' to these reasons---for example, our last two
`bullet-points'---are definitive. But we will from now on
accept that {\em some type\/} of theory of quantum gravity
should be sought.

\subsubsection{Four Types of Approach to Quantum Gravity}
\label{SubSubSec:FourApproachesQG} In seeking such a
theory, there are four broad types of approach one can
adopt. We shall introduce them as answers to a series of
questions; (questions which develop Section
\ref{SubSec:Prologue}'s contrast between the two
strategies, quantisation and emergence). Broadly speaking,
these questions will place them in an order of increasing
radicalism. So, let us ask: should we adopt a diorthotic
scheme in which general relativity is regarded as `just
another classical field theory' to be quantised in a
more-or-less standard way? Or should we instead expect the
theory of quantum gravity to look quite different from
quantized general relativity, but nevertheless have general
relativity emerge from it as some sort of low-energy
(large-length) limit? This option itself breaks down into
two alternatives, according to whether the theory is a
quantization of {\em some\/} classical theory; or something
that is constructed with no prior reference at all to a
classical system. Or---a fourth alternative---should both
quantum theory and general relativity emerge from a theory
that looks quite different from both?

We will now develop the contrast between these four
alternatives. Our survey in later Sections will not need to
decide between them. But of the three programmes that we
discuss in Section \ref{Sec:ResearchProg},
two adopt the first alternative; and indeed, their
implications for the treatment of spacetime are better
understood than those of programmes adopting the other
alternatives. Aspects of the third and fourth alternatives
will be taken up in Section \ref{Sec:TowardsQuST}.

\medskip\noindent
1. {\em Quantise general relativity}. The idea is to start
with the classical theory of general relativity and then to
apply some type of quantisation algorithm. This is intended
to be analogous to the way in which the classical theory
of, for example, an atom bound by the Coulomb potential is
`quantised' by replacing certain classical observables with
self-adjoint operators on a Hilbert space; or, to take
another example, the way in which classical
electromagnetism is quantised to yield quantum
electrodynamics.\footnote{Essentially this approach was
also used in developing important elementary-particle
physics theories, where there was no pre-existing classical
theory; for example, the Salam-Glashow-Weinberg
electro-weak theory, and the quantum chromodynamics
description of the strong nuclear force.}

In the context of quantum gravity, the task is usually
taken to be quantisation of the metric tensor regarded as a
special type of field. In practice, the techniques that
have been adopted fall into two classes: (i) those based on
a spacetime approach to quantum field theory---in which the
operator fields are defined on a four-dimensional manifold
representing spacetime; and (ii) those based on a canonical
approach---in which the operator fields are defined on a
three-dimensional manifold representing physical space. We
shall discuss (i) and (ii) in more detail in Sections
\ref{SubSec:ParticleProg} and \ref{SubSec:CanWQG}
respectively.

\medskip\noindent
2. {\em General relativity as the low-energy limit of a
quantization of a different classical theory}. If a
quantisation algorithm is applied to some classical theory,
then that theory is naturally recovered as a classical
limit of the ensuing quantum theory. In particular, this
procedure provides a natural interpretation of the physical
variables that arise in the quantum theory as the result of
`quantising' the corresponding classical theory.

But there are various senses of `classical limit': it can
refer to special states whose evolution over time follows
classical laws, or to certain quantum quantities taking
values in a range where classical theory is successful. So
given a quantisation of a classical theory, some {\em
other\/} classical theory might also be a classical limit
of it, in some good sense.

Hence the idea in the context of quantum gravity, that
general relativity might emerge as a low-energy (large-distance)
classical limit of a quantum theory, that is given to us as
a quantisation of a different classical theory. Of course,
in view of our lack of data about quantum gravity, one
expects it will be very hard to guess the correct classical
theory from which to start.

But despite this difficulty, this type of approach is
exemplified by the main current research programme:
superstring theory, which quantizes a classical `string
theory' and yet has general relativity as a low-energy
limit. We shall discuss this programme in Section
\ref{SubSec:Superstrings}.\footnote{For more discussion,
see Part 4 of this volume.} For the moment, suffice it to
say that the dimensional nature of the basic Planck units
lends support to the idea of a theory that could reproduce
standard general relativity in regimes whose scales are
well away from that of the Planck time, length, energy,
etc. This remark is reinforced by a well-known body of work
to the effect that, with appropriate caveats, general
relativity is necessarily recovered as the low-energy limit
of {\em any\/} interacting theory of massless spin-2
particles propagating on a Minkowski background, in which
the energy and momentum are conserved \cite{BD75}. The most
notable example of this type is the theory of closed
superstrings which has a natural massless, spin-2
excitation.

However, superstring theory is by no means the only example
of this type of approach. For it is conservative, in that
the classical `string theory' that it quantises assumes the
classical concept of a manifold. Roughly speaking, in
perturbative superstring theory, the quantum variables are
the functions that embed the string in a continuum
spacetime manifold. But there have been more radical
attempts to quantise aspects of space, or spacetime,
itself. For example, there have been several attempts to
construct a quantum theory of topology; and there have been
attempts to quantize causal structures in which the
underlying set is discrete. However, recovering general
relativity as a classical limit of theories of this type is
by no means trivial since the implication is that the {\em
differentiable manifold\/} structure of spacetime, not just
its metric tensor, must be understood in some
phenomenological sense. We will postpone to Section
\ref{Sec:TowardsQuST} discussion of these more radical
attempts to quantise `spacetime itself'.

Finally, a general point about this type of approach. Given
the general scenario where we obtain a classical theory as
a limit of a quantum theory, it is natural to wonder what
would happen if one tried to quantise this derivative
classical theory; (in the case of interest to us, general
relativity). Generally speaking, this does not give back
the initial quantum theory. That is unsurprising, in view
of our comments above about the variety of classical
limits. But we should also make two more specific remarks.
First, one reason why one does not get back the initial
quantum theory may be that the classical limit is
non-renormalisable: this is well known to be the case for
general relativity (as we will discuss in Section
\ref{SubSec:ParticleProg}). But second: this feature does
not render the `re-quantisation' procedure completely
useless. Indeed, genuine quantum predictions can be
obtained by empirically fixing the appropriate number of
renormalisation constants, where what is `appropriate' is
determined by the energy at which the theory is to be
employed. Theories of this type are called ``effective
field theories'' and are a valuable tool in modern
theoretical physics. For a recent review in the context of
general relativity see \cite{Don98}.

\medskip\noindent
3. {\em General relativity as the low-energy limit of a
quantum theory that is not a quantization of a classical
theory}. The procedure of going from classical to quantum
has become so ubiquitous (for example, look at the content
of a typical undergraduate lecture course on quantum
theory!) that one might be tempted to assume that all
quantum theories necessarily arise in this way. However,
there is no good reason why this should be so. So it is
certainly reasonable to consider the construction of a
quantum theory {\em ab initio\/} with no fundamental
reference to an underlying classical theory---for example
as a representation of some group or algebra. The question
then arises whether a quantum theory of this type may have
a classical limit of some sort, even though it is {\em
not\/} obtained by the quantisation of such. A good example
of a quantum theory of this kind was the `current algebra'
approach to strong-interaction physics that was intensely
studied in the 1960s.

Of course, one might well fear that in quantum gravity,
with its dire lack of data, this type of approach will be
at least as hard to implement successfully as is the
previous one: the correct group or algebra might be as hard
to guess as is the correct classical system to quantise in
the previous approach. However, recent developments in
understanding the non-perturbative aspects of superstring
theory suggest that this type of approach may well come to
the fore in that programme; (see Section
\ref{SubSec:Superstrings}).

\medskip\noindent
4. {\em Start ab initio with a radically new theory}. The
idea here is that both classical general relativity {\em
and\/} standard quantum theory emerge from a theory that
looks very different from both. Such a theory would indeed
be radically new. For recall that we classified as examples
of the second type of approach above, quantisations of
spatial or spatiotemporal structure other than the metric;
for example, quantisations of topology or causal structure.
So the kind of theory envisaged here would somehow be still
more radical than that; presumably by not being a quantum
theory, even in a broad sense---for example, in the sense
of states giving amplitudes to the values of quantities,
whose norms squared give probabilities.

Of course, very little is known about potential schemes of
this type; let alone whether it is necessary to adopt such
an iconoclastic position in order to solve the problem of
quantum gravity. We shall mention some possible clues in
Section 5. For the moment, we want just to emphasise the
philosophical interest of this type of approach. For it is
often motivated by the view that the basic ideas behind
general relativity and quantum theory are so fundamentally
incompatible that any complete reconciliation will
necessitate a total rethinking of the central categories of
space, time and matter. And as we mentioned in Section
\ref{SubSec:NoData} (item 3), we like to think that
philosophy could have a role in that enterprise!

\medskip\noindent
As mentioned above, all four types of approach have been
followed in the past (albeit in a very limited way in
regard to the third and fourth types). Until fifteen years ago, the
bulk of the effort was devoted to the first---the active
quantisation of classical general relativity; so that two
of the three programmes reviewed in Section
\ref{Sec:ResearchProg} are of this type. But nowadays the
dominant programme, viz.\ superstrings, is of the second
type; although the second most dominant
programme---canonical quantum gravity in the Ashtekar
approach---is of the first type; and both these programmes
have touches of the third type. In short, it remains a matter of
vigorous debate which of these types of approach will
ultimately prove to be the most fruitful.

\subsection{The Role of Diffeomorphisms}
\label{SubSec:RoleDiffeos} To set the stage for Section
\ref{Sec:ResearchProg}, we devote the rest of this Section
to discussing two conceptual aspects that relate
specifically to spacetime: viz.\ the role of
diffeomorphisms (this Subsection); and the problem of time
(Section \ref{SubSec:ProblemTime}).

\subsubsection{Spacetime Diffeomorphisms in Classical General
Relativity} The group of spacetime diffeomorphisms $\cal D$
plays a key role in classical general relativity; and its
status in quantum gravity raises some major conceptual
issues.\footnote{The diffeomorphisms concerned are those of
compact support, {\em i.e.}, they are equal to the unit map
outside some closed and bounded region of the spacetime
(for some purposes, more subtle `fall-off' rates at
infinity may be appropriate). Thus, for example, a
Poincar\'e-group transformation of Minkowski spacetime is
not included. This restriction is imposed because the role
of transformations with a non-trivial action in the
asymptotic regions of spacetime is quite different from
those that act trivially.}

    In considering these matters, it is important to distinguish
between the pseudo-group of local coordinate
transformations and the genuine group $\cal D$ of global
diffeomorphisms. Compatibility with the former can be taken
to imply that the theory should be written using tensorial
objects on spacetime. On the other hand, diffeomorphisms
are active transformations of spacetime, and invariance
under $\cal D$ implies, we take it, that the points in
spacetime have no direct physical significance (see the
discussion of realism, and the hole argument in Sections
\ref{SubSec:Realism?}, and \ref{SubSec:InterpretGR}.) Of
course, this is also true in special relativity but it is
mitigated there by the existence of inertial reference
frames that can be transformed into each other by the
Poincar\'e group of isometries of the Minkowski metric.

    Put somewhat differently, the action of $\cal D$ induces an
action on the space of spacetime fields, and the only thing
that has immediate physical meaning is the space of
equivalence classes under this action: {\em i.e.}, two
field configurations are regarded as physically equivalent
if they are connected by a diffeomorphism transformation.
Technically, this is analogous in certain respects to the
situation in electromagnetism whereby a vector potential
$A_\mu$ is equivalent to $A_\mu+\partial_\mu f$ for all
functions $f$. However, there is an important difference
between electromagnetism and general relativity.
Electromagnetic gauge transformations occur at a fixed
spacetime point $X$, and the physical configurations can be
identified with the values of the field tensor
$F_{\mu\nu}(X)$, which depends {\em locally\/} on points of
$\cal M$. On the other hand, a diffeomorphism maps one
spacetime point into another, and therefore the obvious way
of constructing a diffeomorphism-invariant object is to
take a scalar function of spacetime fields and integrate it
over the whole of spacetime, which gives something that is
very {\em non\/}-local. The idea that `physical
observables' are naturally non-local is an important
ingredient in some approaches to quantum gravity.

\subsubsection{Diffeomorphisms in Quantum Gravity}
The role of diffeomorphisms in quantum gravity depends
strongly on the approach taken to the subject. For example,
if the structure of classical relativity is expected to
appear only in a low-energy limit---as, for example, is the
case for superstring theory---there is no strong reason to
suppose that the group of spacetime diffeomorphisms $\cal
D$ will play any fundamental role in the quantum theory. On
the other hand, in schemes which involve the active
quantisation of the classical gravitational field, $\cal D$
is likely to be a key ingredient in forcing the quantum
theory to comply with the demands of general relativity.
However, it should be noted that the situation in
`canonical' quantum gravity is less clear-cut: this
programme is based on a prior decomposition into space plus
time, and this is bound to obscure the role of spacetime
diffeomorphisms.

In general terms, there are at least three\footnote{Another
possibility---not exemplified in the programmes surveyed in
this paper---is that the diffeomorphism group $\cal D$
could be related to a bigger group $G$ in a projective way:
{\em i.e.}, there is some normal subgroup $K$ of $G$ so
that $G/K\simeq {\cal D}$.} ways in which $\cal D$ could
appear in the quantum theory; which will be exemplified in
the programmes surveyed in Sections \ref{Sec:ResearchProg}
{\em et seq.\/}:
\begin{enumerate}
\item[(i)] as an {\em exact\/} covariance/invariance
group;\footnote{Here we should distinguish the invariance
of each individual expression in a theory's formalism from
the invariance of all the physically measurable values of
quantities; the latter being of course a weaker property.
In fact, most of the programmes whose basic framework
treats $\cal D$ as an exact covariance group {\em in
practice\/} `fix a gauge' for their formalism, and so work
with non-covariant/non-diffeomorphism-invariant
expressions, and then show physically measurable values to
be gauge-invariant. So in practice, these programmes enjoy
only the weaker property.\label{Footnote:GaugeFix}}

\item[(ii)] as a {\em subgroup\/} of a bigger group;

\item[(iii)] as a limited concept associated with a
phenomenological view of spacetime (or space).
\end{enumerate}

In the first two options one could say that the
diffeomorphisms form a `precise' concept since the
mathematical object that occurs in the formalism is exactly
the classical group $\cal D$. The third option, (iii), is
somewhat different and flows naturally from the view that
spacetime is a phenomenological concept of limited
applicability: the same would then be expected for the
diffeomorphisms of the manifold that models spacetime in
this limited sense. We shall say more about this in Section
\ref{Sec:TowardsQuST}.

The idea that $\cal D$ is an exact covariance/invariance
group (option (i) above) plays a key role in several
approaches to quantum gravity. For example, (i) is one of
the central properties of so-called `topological quantum
field theory', which seems to have potential applications
in quantum gravity. And we will see in Section
\ref{Sec:ResearchProg} that (i) also plays a major role in
the particle-physics programme (Section
\ref{SubSec:ParticleProg}), albeit with the qualification
mentioned in footnote \ref{Footnote:GaugeFix} above [to
option (i)]; and in a less clear-cut way, in canonical
quantum gravity.

On the other hand, the idea that $\cal D$ is a subgroup of
a bigger covariance group (option (ii) above) is endorsed
by the perturbative approach to superstring theory. In
short, the idea is that the extra fields associated with
supersymmetry lead to a much larger covariance group; more
details in Section \ref{SubSec:Superstrings}.

\subsection{The Problem of Time}
\label{SubSec:ProblemTime} Closely related to the role of
diffeomorphisms is the infamous `problem of time'. This problem
is central in any approach to quantum gravity that assigns a
significant {\em prima facie\/} role to classical general
relativity (unlike, say, superstring theory). For the problem
arises from the very different roles played by the concept of
time in quantum theory and in general relativity; and the problem
lies at the heart of many of the deepest conceptual issues in
such approaches to quantum gravity. To present the problem, we
will consider the roles of time, first in quantum theory, and
then in general relativity.\footnote{For much fuller expositions
of the problem from the point of view of canonical quantum
gravity, see the Chapters by Belot \& Earman, and Weinstein, in
Part 2 of this volume; see also \cite{Kuc92a},
\cite{Ish93}.\label{Footnote:ProblemTime}}

\subsubsection{Time in Quantum Theory}
In quantum theory, time is not a physical quantity in the
normal sense, since it is not represented by an operator.
Rather, it is treated as a {\em background\/} parameter
which, as in classical physics, is used to mark the
evolution of the system; witness the parameter $t$ in the
time-dependent Schr\"odinger equation.\footnote{This is why
the meaning assigned to the time-energy uncertainty
relation $\delta t\,\delta E\ge{1\over 2}\hbar$ is quite
different from that associated with, for example, the
position and the momentum of a particle.}

    Besides, the idea of an event happening at a given time plays
a crucial role in the technical and conceptual foundations
of quantum theory:
 \begin{itemize}
     \item One of the central requirements of the
{\em scalar product\/} on the Hilbert space of states is
that it is conserved under the time evolution given by the
Schr\"odinger equation. This is closely connected to the
unitarity requirement that probabilities always sum to one.

        \item More generally, a key ingredient in the construction of
the Hilbert space for a quantum system is the selection of
a complete set of quantities that are required to commute
at a fixed value of time.

        \item Conceptually, the notion of {\em measuring\/} a quantity
at a given time, to find its value at that time, is a
fundamental ingredient of both the minimal statistical
interpretation of the theory, and the Copenhagen
interpretation (see Section
\ref{SubSubSec:Instrumentalism}).
\end{itemize}

    Furthermore, all these ideas can be extended to systems that are
compatible with special relativity: the unique time system
of Newtonian physics is simply replaced with the set of
relativistic inertial reference frames. The quantum theory
can be made independent of a choice of frame, provided that
the theory carries a unitary representation of the
Poincar\'e group of isometries of the metric of Minkowski
spacetime. In the case of a relativistic quantum field
theory, the existence of such a representation is closely
related to the microcausality requirement that fields
evaluated at spacelike-separated points must commute. For
example, a scalar quantum field $\widehat\phi(X)$ satisfies
the commutation relations
\begin{equation}
[\,\widehat\phi(X),\,\widehat\phi(Y)\,]=0
\label{microcausality}
\end{equation}
whenever the spacetime points $X$ and $Y$ are spacelike
separated.

    Finally, we note that this background time is truly an abstraction
in the sense that according to quantum theory, no {\em
physical\/} clock can provide a precise measure of it
\cite{UW89}: there is always a small probability that a
real clock will sometimes run backwards with respect to it.

\subsubsection{Time in General Relativity; and the Problem of Time}
\label{SubSubSec:TimeGRProbTime} When we turn to classical
general relativity, the treatment of time is very
different. Time is not treated as a background parameter,
even in the liberal sense used in special relativity, viz.\
as an aspect of a fixed, background spacetime structure.
Rather, what counts as a choice of a time ({\em i.e\/} of a
timelike direction) is influenced by what matter is
present; (as is, of course, the spatial metrical
structure). The existence of many such times is reflected
in the fact that if the spacetime manifold has a topology
that enables it to be foliated as a one-parameter family of
spacelike surfaces, this can generally be done in many
ways---without any subset of foliations being singled out
in the way families of inertial reference frames are
singled out in special relativity. From one perspective,
each such parameter might be regarded as a legitimate
definition of (global) time. However, in general, there is
no way of selecting a particular foliation, or a special
family of such, that is `natural' within the context of the
theory alone. In particular, these definitions of time are
in general unphysical, in that they provide no hint as to
how their time might be measured or registered.

But the main problem about time in general relativity
arises when we turn to quantum gravity, where the disparate
nature of the treatments of time in quantum theory and in
general relativity becomes of paramount significance. We
shall see various more specific versions of this problem in
each of the research programmes reviewed in Section
\ref{Sec:ResearchProg}. But for the moment, we introduce
the problem in general terms.

General relativity accustoms us to the ideas that: (i) the
causal structure of spacetime depends on the metric
structure, represented by the metric tensor $\gamma$; and
(ii) the metric and causal structures are influenced by
matter, and so vary from one model of the theory to
another. In general relativity, these ideas are `kept under
control' in the sense that in each model, there is of
course a {\em single\/} metric tensor $\gamma$,
representing a single metric and causal structure. But once
we embark on constructing a quantum theory of gravity, we
expect some sort of quantum fluctuations in the metric, and
so also in the causal structure. But in that case, how are
we to formulate a quantum theory with a fluctuating causal
structure?

This general statement of the problem is clearly relevant
if one proposes a spacetime-oriented approach to
formulating the quantum theory; since then one's prototype
quantum theories will emphasise a fixed background causal
structure. But the same statement of the problem arises on
various other approaches to quantum gravity. For example,
if one takes the view that the spacetime metric is only a
coarse-grained, phenomenological construct of some type,
then so is the causal structure. And again, the question
arises how we are to formulate a quantum theory with such a
causal structure.\footnote{Note that this general statement
of the problem of time would have an analogue in a
stochastic version of {\em classical\/} general relativity,
in which the metric tensor is regarded as a random
variable. Of course, the other difficult problem of
understanding what is meant by `superpositions' of
spacetime geometries would be absent in this case.}

Though this problem is at bottom conceptual, it has clear
technical aspects. In particular, a probabilistic causal
structure poses severe technical problems for relativistic
quantum field theory, whose standard formulation
presupposes a fixed causal structure. For example, a
quantum scalar field satisfies the microcausal commutation
relations in Eq.\ (\ref{microcausality}), whereby fields
evaluated at spacelike separated spacetime points commute.
However, the concept of two points being spacelike
separated has no meaning if the spacetime metric is
probabilistic or phenomenological. In the former case, the
most likely scenario is that the right hand side of the
commutator in Eq.\ (\ref{microcausality}) never vanishes,
thereby removing one of the foundations of conventional
quantum field theory.

In practice, the techniques that have been used to address the
problem of time fall into one of the following three strategies,
to all of which we shall return in Section
\ref{Sec:ResearchProg}:
\begin{enumerate}
\item Use a fixed background metric---often
chosen to be that of Minkowski spacetime---to define a
fiducial causal structure with respect to which standard
quantum field theoretical techniques can be employed. This
is the strategy adopted by the old particle-physics
programme, using Minkowski spacetime (Section
\ref{SubSec:ParticleProg}).

This strategy raises questions about how the background
structure is to be chosen. Of course, Minkowski spacetime
seems very natural from the perspective of standard quantum
field theory, but it is rather arbitrary when seen in the
context of general relativity. One possibility is that the
background structure could come from a {\em contingent\/}
feature of the actual universe; for example, the $3^0$K
microwave background radiation. However, structure of this
type is approximate and is therefore applicable only if
fine details are ignored. In addition, the problem of
rigorously constructing (even free) quantum fields has only
been solved for a very small number of background
spacetimes; certainly there is no reason to suppose that
well-defined quantum field theories exist on a generic
spacetime manifold. Also, there is a general matter of
principle: should we require a theory of quantum gravity to
work for `all possible' universes (however that is made
precise), or can it depend on special features of the
actual one in which we live?

\item Accept the fact that there is no background spacetime
reference system and attempt to locate events, both
spatially and temporally, with specific functionals of the
gravitational and other fields. This important idea is of
course motivated by the analysis of the `hole argument' and
spacetime diffeomorphisms (see Sections
\ref{SubSec:InterpretGR} and \ref{SubSec:RoleDiffeos}).
Thus the idea is that for the example of a scalar field
$\phi$, the value $\phi(X)$ of $\phi$ at a particular
spacetime point $X$ has no physical meaning because of the
action of the spacetime diffeomorphism group; but, the
value of $\phi$ where something `is' {\em does\/} have a
physical meaning in the sense that `$\phi({\rm thing})$' is
diffeomorphism invariant. In practice, this strategy seems
only to have been adopted by some of the
approaches to the problem of time as it manifests itself in
canonical quantum gravity.

\item {And indeed, one approach is to drop spacetime methods
and instead adopt a canonical approach to general
relativity, so that the basic ingredients are geometrical
fields on a three-dimensional manifold. The problem then is
to reconstruct some type of spatio-temporal picture within
which the quantum calculations can be interpreted. This is
the strategy adopted by the canonical quantum gravity
programme (Section \ref{SubSec:CanWQG}).

Studies of the problem of time in canonical quantum gravity
raise the alluring question whether a meaningful quantum
theory can be constructed in a way that contains no
fundamental reference to time at all. That this is a far
from trivial matter is shown by our earlier remarks about
the crucial role of time in conventional quantum theory
(see the references in footnote
\ref{Footnote:ProblemTime}).}
\end{enumerate}

\section{Research Programmes in Quantum Gravity}
\label{Sec:ResearchProg} As we have seen, we are far from
having an `axiomatic' framework for quantum gravity, or
even a broad consensus about what to strive for beyond the
minimal requirement that the theory should reproduce
classical general relativity and normal quantum theory in
the appropriate domains---usually taken to be all physical
regimes well away from those characterised by the Planck
length.

In this Section, we shall focus on three specific research
programmes. Our aim is not to review the technical status
of these programmes, but rather to explore their treatments
of spacetime. Of these three programmes, two are the main
current focus for work in quantum gravity: superstring
theory, and canonical quantum gravity (in the version
called `the Ashtekar programme'). These programmes
complement each other nicely, and enable the special ideas
of either of them to be viewed in a different perspective
by invoking the other---a feature that is rather useful in
a subject that lacks unequivocal experimental data! For
example, they exemplify the choice of approaches we
discussed in Section \ref{SubSubSec:FourApproachesQG},
about whether to quantise general relativity, or to have
classical general relativity emerge from a quantum theory
of something quite different. Superstring theory takes the
latter approach; canonical quantum gravity the former.

The other programme we shall discuss is a
spacetime-oriented quantization of general relativity,
which we dub `the particle-physics programme'. This
programme is no longer regarded as capable of providing a
full theory of quantum gravity; but it predated, and so
influenced, both the other two programmes, and this means
that discussing its own treatment of spacetime will form a
helpful backdrop to discussing theirs.\footnote{Clearly,
one could also discuss other programmes so as to provide
still more of a backdrop to the two main ones. One obvious
choice is the Euclidean programme, which can be viewed as a
spacetime-oriented species of the traditional
(geometrodynamic, rather than Ashtekar) version of
canonical quantum gravity. But we ignore it here, since (i)
it is especially connected with quantum cosmology, which we
have set aside; and (ii) we discuss it in \cite{BI99a}.}

All three programmes postulate at the fundamental level a
spacetime manifold. But it may differ in its dimension,
metric structure etc.\ from the 4-dimensional Lorentzian
manifold familiar from classical general relativity. And,
in fact, these programmes suggest limitations to the
applicability of the concept of a spacetime manifold
itself. We shall explore this possibility further in
Section \ref{Sec:TowardsQuST}.

We begin in Section \ref{SubSub:FocusQuestion} by listing
four topics that will act as `probes' in our survey of how
concepts of space and time are treated in these programmes.
This is followed by some historical orientation to these
programmes (Section \ref{SubSec:History}), and then we
consider them {\em seriatim\/}, in the following order: the
particle-physics programme (Section
\ref{SubSec:ParticleProg}); supergravity and superstrings
(Section \ref{SubSec:Superstrings}); and canonical quantum
gravity (Section \ref{SubSec:CanWQG}).

\subsection{Focussing the Question: How is Spacetime Treated?}
\label{SubSub:FocusQuestion} We will take the following
four topics as `probes' in our survey of how the concept of
spacetime is treated in the various quantum
gravity programmes. We present them as a sequence of
questions; but of course they overlap with one another.

\begin{enumerate}

    \item { {\em Use of standard quantum theory\/}. Are the
technical formalism and conceptual framework of present-day
quantum theory adequate for the programme's envisaged
theory of quantum gravity? In particular, do any features
of the programme suggest advantages, or indeed
disadvantages, of the `heterodox' interpretations of
quantum theory, discussed in Sections
\ref{SubSubSec:Literalism}--\ref{SubSubSec:NewDynamics}?}

    \item { {\em Use of standard spacetime concepts\/}. How much
of the familiar treatment of spatio-temporal concepts,
adopted by general relativity and quantum theory, does the
programme adopt? In particular, does it model spacetime as
a 4-dimensional differentiable manifold? If so, does it add
to this manifold a quantized metric tensor? And if so, what
exactly is the relation of this to a classical Lorentzian
metric on the manifold?}

    \item{ {\em The spacetime diffeomorphism group\/}. Assuming
the programme models spacetime as a manifold, what role
does it assign to the group of spacetime diffeomorphisms?
Or if it decomposes spacetime into space and time, what
role is given to spatial diffeomorphisms? }

\item { {\em The problem of time\/}. How does the problem of time
manifest itself in the programme, and how does the
programme address it? In particular: how much of the
familiar treatment of spacetime must be retained for the
envisaged theory of quantum gravity to be constructed? Must
both metric and manifold be fixed; or can we work with a
fixed background manifold, but no background metric? Could
we also do without the background manifold?}

\end{enumerate}

\subsection{Some Historical Background to the Three Programmes}
\label{SubSec:History} To introduce the survey of our three
chosen research programmes, it is useful to sketch some of
the historical development of quantum gravity research---a
development from which these programmes have all sprung.

The early history of attempts to quantise general
relativity goes back at least to the 1960s and was marked
by a deep division of opinion about whether quantum gravity
should be tackled from a spacetime perspective---the
so-called `covariant' approach, whose leading champion was
Bryce DeWitt---or from a `canonical' approach, in which
spacetime is decomposed into space plus time before the
theory is quantised.

The early predominance of the canonical programme stemmed
partly from the fact that the attitude in the 1960s towards
quantum field theory was very different from that of today.
With the exception of quantum electrodynamics, quantum
field theory was poorly rated as a fundamental way of
describing the interactions of elementary particles.
Instead, this was the era of the S-matrix, the Chew axioms,
Regge poles, and---towards the end of that period---the
dual resonance model and the Veneziano amplitude that led
eventually to superstring theory.

    In so far as it was invoked at all in strong interaction
physics, quantum field theory was mainly used as a
phenomenological tool to explore the predictions of current
algebra, which was thought to be more fundamental. When
quantum field theory was studied seriously, it was largely
in the context of an `axiomatic' programme---such as the
Wightman axioms for the $n$-point functions.

    This neglect of quantum field theory influenced the way
quantum gravity developed. In particular---with a few
notable exceptions---physicists trained in particle physics
and quantum field theory were not interested in quantum
gravity, and the subject was mainly left to those whose
primary training had been in general relativity. This
imparted a special flavour to much of the work in that era.
In particular, the geometrical aspects of the theory were
often emphasised at the expense of quantum field theoretic
issues---thereby giving rise to a tension that has affected
the subject to this day.

    However, a major change took place in the early 1970s when
t'Hooft demonstrated the renormalisability of quantised
Yang-Mills theory. Although not directly connected with
gravity, these results had a strong effect on attitudes
towards quantum field theory in general, and reawakened a
wide interest in the subject. One spin-off was that many
young workers in particle physics became intrigued by the
challenge of applying the new methods to quantum gravity.
This led to a revival of the covariant approach; more
specifically, to the particle-physics programme (Section
\ref{SubSec:ParticleProg}), and thereby eventually to
supergravity and superstring theory (Section
\ref{SubSec:Superstrings}).

    On the other hand, the canonical programme has also continued
to flourish since the 1970s; indeed, until the relatively
recent advent of `superstring cosmology', canonical quantum
gravity provided the only technical framework in which to
discuss quantum cosmology. A major development in canonical
quantum gravity was Ashtekar's discovery in 1986 of a new
set of variables that dramatically simplifies the
intractable Wheeler-DeWitt equation which lies at the heart
of the programme's quantum formalism.

\subsection{The Particle-Physics Programme}
\label{SubSec:ParticleProg}
\subsubsection{The basic ideas}
In this programme, the basic entity is the {\em
graviton\/}---the quantum of the gravitational field. Such
a particle is deemed to propagate in a background Minkowski
spacetime, and---like all elementary particles---is
associated with a specific representation of the Poincar\'e
group which is labelled by its mass and its spin. The
possible values of mass and spin are sharply limited by the
physical functions which the graviton is to serve. In
particular, replication of the inverse-square law behaviour
of the static gravitational force requires the graviton to
have mass zero, and the spin must be either $0$ or
$2\hbar$. However, zero spin is associated with a scalar
field $\phi(X)$, whereas spin-two comes from a symmetric
Lorentz tensor field $h_{\mu\nu}(X)$; the obvious
implication is that these spin values correspond to
Newtonian gravity and general relativity respectively.

The key to relating spin-2 particles and general relativity
is to fix the background topology and differential
structure of spacetime $\cal M$ to be that of Minkowski
spacetime, and then to write the Lorentzian metric $\gamma$
on $\cal M$ as
\begin{equation}
\gamma_{\alpha\beta}(X)=\eta_{\alpha\beta}+\kappa
h_{\alpha\beta}(X).
            \label{g=eta+h}
\end{equation}
Here, $h$ measures the departure of $\gamma$ from the flat
spacetime metric $\eta$ and is regarded as the `physical'
gravitational field with the coupling constant
$\kappa^2=8\pi G/c^2$ where $G$ is Newton's constant.

The use of the expansion in Eq.\ (\ref{g=eta+h}) strongly
suggests a perturbative approach in which quantum gravity
is seen as a theory of small quantum fluctuations around a
background Minkowski spacetime. Indeed, when this expansion
is substituted into the Einstein-Hilbert action for pure
gravity, $S=\int d^4X |\gamma|^{\frac12} R(\gamma)$ (where
$R(\gamma)$ is the scalar curvature), it yields (i) a term
that is bilinear in the fields $h$ and which---when
quantised in a standard way---gives a theory of
non-interacting, massless spin-2 gravitons; and (ii) a
series of higher-order terms that describe the interactions
of the gravitons with each other. Thus a typical task would
be to compute the probabilities for various numbers of
gravitons to scatter with each other and with the quanta of
whatever matter fields might be added to the system.

    This approach to quantum gravity has some problematic
conceptual features (see below). But, nonetheless, had it
worked it would have been a major result, and would
undoubtedly have triggered a substantial effort to
construct a spacetime-focussed quantum gravity theory in a
non-perturbative way. A good analogue is the great increase
in studies of lattice gauge theory that followed the proof
by t'Hooft that Yang-Mills theory is perturbatively
renormalisable.

However, this is not what happened. Instead, a number of
calculations were performed around 1973 that confirmed
earlier suspicions that perturbative quantum gravity is
non-renormalisable\footnote{Full references can be found in
reviews written around that time; for example, in the
proceedings of the first two Oxford conferences on quantum
gravity \cite{IPS75,IPS81}.}. There have been four main
reactions to this situation:
\begin{itemize}

\item Adopt the view in which general relativity is an
``effective field theory'' and simply add as many
empirically determined counter-terms as are appropriate at
the energy concerned. The ensuing structure will break down
at the Planck scale but a pragmatic particle physicist
might argue that this is of no importance since the Planck
energy is so much larger than anything that could be
feasibly attainable in any foreseeable particle accelerator
(see the discussion at the end of the second approach in
Section \ref{SubSubSec:FourApproachesQG}.)

    \item Continue to use standard perturbative quantum field
theory but change the classical theory of general
relativity so that the quantum theory becomes
renormalisable. Examples of such attempts include (i)
adding higher powers of the Riemann curvature
$R^\alpha_{\beta\mu\nu}(\gamma)$ to the action; and (ii)
supergravity (see Section \ref{SubSec:Superstrings} below).

    \item Keep classical general relativity as it is, but develop
quantisation methods that are intrinsically
non-perturbative. Examples of this philosophy are `Regge
calculus' (which involves simplicial approximations to
spacetime) and techniques based on lattice gauge theory. Of
particular importance in recent years is the Ashtekar
programme for canonical quantisation which is fundamentally
non-perturbative (see Section \ref{SubSubSec:Ashtekar} below).

    \item Adopt the view that the non-renormalisability of
perturbative quantum gravity is a catastrophic failure that
requires a very different type of approach. In terms of the
classification in Section \ref{SubSubSec:FourApproachesQG},
this would mean adopting its second, or third or fourth
types of approach: quantising a classical theory that is
quite different from general relativity (such as a string
theory); or having general relativity emerge as a
low-energy limit of a quantum theory that is not a
quantisation of any classical system; or having it and
quantum theory both emerge from something completely
different.

\end{itemize}

\subsubsection{Spacetime according to the
Particle-Physics Programme} The response given by this
programme to our four conceptual probes presented in
Section \ref{SubSub:FocusQuestion} is as follows.

\smallskip\noindent
1. {\em Use of standard quantum theory\/}. The basic
technical ideas of standard quantum theory are employed,
suitably adapted to handle the gauge structure of the
theory of massless spin-2 particles. Furthermore, the
traditional, Copenhagen interpretation of the theory is
applicable (even if not right!), in that the background
Minkowski metric and spacetime manifold are available to
serve as the classical framework, in which measurements of
the quantum system, according to this interpretation, are
to be made.

Of course, other interpretations of quantum theory
discussed in Section \ref{SubSec:InterpretQT}, such as
Everettian or pilot-wave interpretations, may well also be
applicable to this programme. Our present point is simply
that the particle-physics programme gives no special
reasons in favour of such interpretations.

\smallskip\noindent
2. {\em Use of standard spacetime concepts\/}. The
background manifold and metric are described in the
language of standard differential geometry. Note that, from
a physical perspective, the restriction to a specific
background topology means a scheme of this type is not well
adapted to addressing some of the most interesting
questions in quantum gravity such as black-hole phenomena,
quantum cosmology, the idea of possible spacetime `phase
changes' etc.

\smallskip\noindent
3. {\em The spacetime diffeomorphism group\/}. The action
of the group of spacetime diffeomorphisms is usually
studied infinitesimally. The transformation of the graviton
field $h_{\mu\nu}(X)$ under a vector-field generator $\xi$
of such a diffeomorphism is simply $h_{\mu\nu}(X)\mapsto
h_{\mu\nu}(X)+\partial_\mu\xi_\nu(X)+\partial_\nu\xi_\mu(X)$,
which is very reminiscent of the gauge transformations of
the electromagnetic vector potential, $A_\mu(X)\mapsto
A_\mu(X)+\partial_\mu\phi(X)$.

Indeed, in this infinitesimal sense, the effect of
spacetime diffeomorphisms is strictly analogous to the
conventional gauge transformations of electromagnetism or
Yang-Mills theory (in that spacetime points are fixed); and
the same type of quantisation procedure can be used. In
particular, the invariance of the quantum theory under
these transformations is reflected in a set of `Ward
identities' that must be satisfied by the vacuum
expectation values of time-ordered products (the `$n$-point
functions') of the operator field at different spacetime
points.

\smallskip\noindent
4. {\em The problem of time} The background metric $\eta$
provides a fixed causal structure with the associated
family of Lorentzian inertial frames. Thus, at this level,
there is no problem of time. The causal structure also
allows a notion of microcausality, thereby permitting a
conventional type of relativistic quantum field theory to
be applied to the field $h_{\alpha\beta}$.

However, many people object strongly to an expansion like
Eq.\ (\ref{g=eta+h}) since it is unclear how this
background causal structure is to be related to the
physical one; or, indeed, what the latter really means. For
example, does the `physical' causal structure depend on the
state of the quantum system? There have certainly been
conjectures to the effect that a non-perturbative
quantisation of this system would lead to quantum
fluctuations of the causal structure around a
quantum-averaged background that is not the original
Minkowskian metric. As emphasised earlier, it is not clear
what happens to the microcausal commutativity condition in
such circumstances; or, indeed, what is meant in general by
`causality' and `time' in a system whose light-cones are
themselves the subject of quantum fluctuations.

\subsection{The Superstrings Programme}
\label{SubSec:Superstrings}
\subsubsection{The introduction of supersymmetry}
When confronted with the non-renormalisability of covariant
quantum gravity, the majority of particle physicists
followed a line motivated by the successful transition from
the old non-renormalisable theory of the weak interactions
(the `four-fermion' theory) to the new renormalisable
unification of the weak and electromagnetic forces found by
Salam, Glashow and Weinberg. Thus the aim was to construct
a well-defined theory of quantum gravity by adding
carefully chosen matter fields to the classical theory of
general relativity with the hope that the ultraviolet
divergences would cancel, leaving a theory that is
perturbatively well-behaved.

A key observation in this respect is that the divergence
associated with a loop of gravitons might possibly be
cancelled by introducing {\em fermions\/}, on the grounds
that the numerical sign of a loop of virtual fermions is
opposite to that of a loop of bosons. With this motivation,
supergravity was born, the underlying supersymmetry
invariance being associated with a spin-${3/2}$ fermionic
partner (the `gravitino') for the bosonic spin-$2$
graviton. Moreover, since supersymmetry requires very
special types of matter, such a scheme lends credence to
the claim that a successful theory of quantum gravity must
involve unifying the fundamental forces; {\em i.e.}, the
extra fields needed to cancel the graviton infinities might
be precisely those associated with some grand unified
scheme.

Recall from Section \ref{SubSec:ApproachesQG} that the mark
of unification of two forces is the equality of their
coupling constants; and that the energy-dependence of the
coupling constants for the electromagnetic, weak and strong
nuclear forces renders them at least approximately equal at
around $10^{20}$Mev. Here, it is of considerable interest
to note that the introduction of supersymmetry greatly
improves the prospects for exact equality: there is every
expectation that these three fundamental forces do unify
exactly in the supersymmetric version of the theory. The
fact that $10^{20}$Mev is not that far away from the Planck
energy ($10^{22}$Mev) adds further weight to the idea that
a supersymmetric version of gravity may be needed to
guarantee its inclusion in the pantheon of unified forces.

    Early expectations for supergravity theory were high
following successful low-order results, but it is now
generally accepted that if higher-loop calculations could
be performed (they are very complex) intractable
divergences would appear once more. However, this line of
thought continues and the torch is currently carried by the
superstring programme, which in terms of numbers of papers
produced per week is now by far the dominant research
programme in quantum gravity.

 The superstrings programme has had two phases.  The first phase
began in earnest in the mid-1980s (following seminal work
in the mid 1970s) and used a perturbative approach; as we
shall see, its treatment of spacetime can be presented
readily enough in terms of our four probes listed in
Section \ref{SubSub:FocusQuestion}. The second phase began
in the early 1990s, and `still rages'. It has yielded rich
insights into the underlying non-perturbative theory. But
the dust has by no means settled. Even the overall
structure of the underlying theory remains very unclear;
and in particular, its treatment of spacetime is too
uncertain for our four probes to be applied. Accordingly,
our discussion of the second phase will forsake the probes,
and just report how recent developments indicate some
fundamental limitations in the manifold conception of
spacetime.\footnote{For more detailed discussion of
superstrings, see Part 4 of this volume.}

\subsubsection{Perturbative superstrings}
The perturbative superstrings programme involves quantising a
classical system; but the system concerned is not general
relativity, but rather a system in which a one-dimensional closed
string propagates in a spacetime $\cal M$ (whose dimension is in
general not 4). More precisely, the propagation of the string is
viewed as a map $X:{\cal W}\rightarrow {\cal M}$ from a
two-dimensional `world-sheet' $\cal W$ to spacetime $\cal M$ (the
`target spacetime'). The quantisation procedure quantises $X$,
but not the metric $\gamma$ on $\cal M$, which remains classical.
The appropriate classical theory for the simplest such system is
described by the famous Polyakov action; which is invariant under
conformal transformations on $\cal W$. To preserve this conformal
invariance in the quantised theory, and to satisfy other
desirable conditions, the following conditions are necessary:
\begin{enumerate}
\item the theory is made supersymmetric;

\item the spacetime $\cal M$ has a certain critical dimension
(the exact value depends on what other fields are added to
the simple bosonic string); and

\item the classical spacetime metric
$\gamma$ on $\cal M$ satisfies a set of field equations that are
equivalent to the (supergravity version of) Einstein's field
equations for general relativity plus small corrections of Planck
size: this is the sense in which general relativity emerges from
string theory as a low-energy limit.\footnote{The more
realistic superstring theories involve an additional massless
`dilaton' scalar field $\phi$, and a massless vector particle
described by a three-component field strength $H_{\mu\nu\rho}$.
The presence of these extra fundamental fields has a major effect
on the classical solutions of the field equations; in particular,
there have been many studies recently of black-hole and
cosmological solutions.}
\end{enumerate}

    Superstring theory has the great advantage over the
particle-physics programme that, for certain string theories, the
individual terms in the appropriate perturbation expansion
{\em are\/} finite. Furthermore, the particle content of
theories of this type could be such as to relate the
fundamental forces in a unified way. Thus these theories
provide a concrete realisation of the old hope that quantum
gravity necessarily involves a unification with the other
fundamental forces in nature.

\subsubsection{Spacetime according to the Perturbative
Superstrings Programme} The response given by the
perturbative superstrings programme to the four conceptual
probes in Section \ref{SubSub:FocusQuestion} is broadly as
follows.

\smallskip\noindent
1. {\em Use of standard quantum theory\/}. As in the case
of the particle-physics programme, the basic technical
ideas of standard quantum theory are employed, albeit with
suitable adaptations to handle the gauge structure of the
theory. Similarly, the Copenhagen interpretation of the
theory is, arguably, applicable; namely, by using as the
required classical framework or background structure, the
solution of the low-energy field equations for the
spacetime metric $\gamma$. On the other hand, one can also
argue contrariwise, that it is unsatisfactory to have the
interpretation of our fundamental quantum theory only apply
in a special regime, viz.\ low energies.

 And as for the particle-physics programme, other
interpretations of quantum theory discussed in Section
\ref{SubSec:InterpretQT} might be applicable. But they have
not been developed in connection with superstrings, and so
far as we can see, perturbative superstring theory gives no
special reason in favour of them.

\smallskip\noindent
2. {\em Use of standard spacetime concepts\/}. In
perturbative superstring theories, the target spacetime
$\cal M$ is modelled using standard differential geometry,
and there seems to be no room for any deviation from the
classical view of spacetime. However, in so far as the
dimension of $\cal M$ is greater than four, some type of
`Kaluza-Klein' scenario is required in which the extra
dimensions are sufficiently curled up to produce no
perceivable effect in normal physics, whose arena is a
$4$-dimensional spacetime.

\smallskip\noindent
3. {\em The spacetime diffeomorphism group\/}. Superstring
theory shows clearly how general relativity can occur as a
fragment of a much larger structure---thereby removing much
of the fundamental significance formerly ascribed to the
notions of space and time. True, the low-energy limit of
these theories is a form of supergravity but, nevertheless,
standard spacetime ideas do not play a central role. This
is reflected by the graviton being only one of an infinite
number of particles in the theory. In particular, the
spacetime diffeomorphism group $\cal D$ appears only as
part of a much bigger structure, as in option (ii) of the
discussion in Section \ref{SubSec:RoleDiffeos}.
Consequently, its technical importance for the quantum
scheme is largely subsumed by the bigger group.

\smallskip\noindent
4. {\em The problem of time} The perturbative expansion in
a superstring theory takes place around the background
given by the solution to the low-energy field equations for
the spacetime metric $\gamma$. In particular, this provides
a background causal structure; and hence---in that
sense---there is no problem of time.

On the other hand, the situation is similar in many
respects to that of the particle-physics programme; in
particular, there are the same worries about the meaning of
`causality' and `time' in any precise sense. There is,
however, an important difference between superstring theory
and the simple perturbative quantisation of general
relativity---namely, there are realistic hopes that a
proper non-perturbative version of the former is
technically viable. If such a theory is found it will be
possible to address the conceptual problems with time and
causality in a direct way---something that is very
difficult in the context of the (mathematically
non-existent!) theory of perturbatively quantised general
relativity.

\subsubsection{The Second Phase of Superstrings}
 Since the early 1990s, a lot of work
in the superstrings programme has focussed on exploring the
underlying {\em non\/}-perturbative theory. These
developments seem to have striking implications for our
conception of space and time at the Planck scale. So now we
turn to these, albeit with trepidation, since the dust
has by no means yet settled in this area of research; and
{\em what\/} the implications are is far from
clear!

These developments are based on various types of `duality'
transformation or symmetry. For example, one of the
simplest forms of duality (`$T$-duality') arises when the
target space is a five-dimensional manifold of the form
${\cal M}_4\times S^1$; ($S^1$ is the circle). It
transpires that the physical predictions of the theory are
invariant under replacement of the radius $R$ of the fifth
dimension with $2\alpha'/R$. Thus we cannot differentiate
physically between a very small, and a very large, radius
for the additional dimension---indeed, there is a precise
sense in which they are `gauge' equivalent to each other.
One of the most important implications of this invariance
is that there exists a minimum length of $R_{{\rm
min}}=\sqrt2\alpha'$,\footnote{The phenomenon can be
generalised to more than one extra dimension and with a
topology that is more complex than just a product of
circles.}---an idea that must surely have significant
implications for our overall understanding of the
conceptual implications of the theory.

    Another type of duality (`$S$-duality') involves the
idea that, for certain theories the physics in the
large-coupling limit is given by the weak-coupling limit of
a `dual' theory whose fundamental entities can be
identified with solitonic excitations of the original
theory. It is believed that the several known consistent
perturbative superstring theories are related in this way
to each other\footnote{A third type of duality known as
`mirror symmetry' also plays an important role here.} and
also to the theories involving extended objects
(`membranes') of dimension greater than one. Ideas of this
type are certainly attractive, not least because they
provide a real possibility of theoretically probing the
physically interesting, high-energy regimes of such
theories.

In short, these developments suggest rather strongly that
the manifold conception of spacetime is not applicable at
the Planck length; but is only an emergent notion,
approximately valid at much larger length scales. We shall
take up this idea, in general terms, in Section
\ref{Sec:TowardsQuST}. At a more technical level, the new
ideas suggest that Lagrangian field-theoretic methods
(which are used in the perturbative superstring theories)
are reaching the limit of their domain of applicability,
and should be replaced by---for example---a more algebraic
approach to theory construction that places less reliance
on an underlying classical system of fields; ({\em i.e.\/},
the third of the four types of approach we listed in
Section \ref{SubSubSec:FourApproachesQG}).

\subsection{The Canonical Quantum Gravity Programme}
\label{SubSec:CanWQG}
\subsubsection{Quantum Geometrodynamics}
The canonical approach to quantum gravity starts with a
reference foliation of spacetime with respect to which the
appropriate canonical variables are defined.\footnote{A
fairly comprehensive bibliography of papers on canonical
general relativity can be found in the review papers
\cite{Ish92,Kuc93}.} These are the
$3$-metric $g_{ab}(x)$ on a spatial slice $\Sigma$ of the
foliation, and a canonical conjugate $p^{ab}(x)$
that---from a spacetime perspective---is related to the
extrinsic curvature of $\Sigma$ as embedded in the
four-dimensional spacetime.

    A key property of general relativity---which reflects the
role of the group $\cal D$ of spacetime
diffeomorphisms---is that these variables are not
independent, but instead satisfy certain constraints,
usually written as
\begin{eqnarray}
    {\cal H}_{a}(x)&=& 0 \label{Ha=0}        \\
    {\cal H}_\perp(x)&=& 0          \label{Hperp=0}
\end{eqnarray}
where ${\cal H}_a(x)$ and ${\cal H}_\perp(x)$ are
complicated functions of the $g$ and $p$, and their
derivatives.

    The constraint functions ${\cal H}_a$ and ${\cal H}_\perp$
play a fundamental role in the theory since their Poisson
bracket algebra (known as the `Dirac algebra') is that of
the group $\cal D$ of spacetime diffeomorphisms projected
along, and normal to, the spacelike hypersurfaces of the
reference foliation. Thus the basic question of
understanding the role of spacetime diffeomorphisms is
coded in the structure of these constraints.

    In addition to the constraint equations Eqs.\
(\ref{Ha=0}--\ref{Hperp=0}), there is also a collection of
dynamical equations that specify how the canonical fields
$g_{ab}(x)$ and $p^{cd}(x)$ evolve with respect to the time
variable associated with the given foliation. However, it
transpires that these equations are redundant since it can be
shown that if $\gamma$ is a spacetime metric on $\cal M$ that
satisfies the constraint equations Eqs.\
(\ref{Ha=0}--\ref{Hperp=0}) on any spacelike hypersurface, then
{\em necessarily\/} the projected canonical variables $g_{ab}(x)$
and $p^{cd}(x)$ will satisfy the dynamical equations. In this
sense, the entire theory is already coded into just the four
constraint equations Eqs.\ (\ref{Ha=0}--\ref{Hperp=0}); so, in
practice, attention is almost invariably focussed on them alone.
Furthermore, among these equations, Eq.\ (\ref{Hperp=0}) is the
crucial one, essentially because---when viewed from a spacetime
perspective---${\cal H}_\perp$ is associated with the canonical
generators of displacements in time-like directions.

    This system can be quantised in a variety of ways.  One
possibility is to impose a gauge for the invariance
associated with the Dirac algebra; solve the constraints
Eqs.\ (\ref{Ha=0}--\ref{Hperp=0}) classically; and then
quantise the resulting `true' canonical system in a
standard way. However, the final equations are intractable
in anything other than a perturbative sense, where they
promptly succumb to virulent ultraviolet divergences.

    Most approaches to canonical quantum gravity do not
proceed in this way. Instead, the full set of fields
$(g_{ab}(x), p^{cd}(x))$ is quantised via the `canonical
commutation relations'
\begin{eqnarray}
    [\,\widehat g_{ab}(x),\widehat g_{cd}(x')\,]  &=& 0
                                                \label{CR:gg}\\
    {[}\,\widehat p^{ab}(x),\widehat p^{cd}(x')\,] &=& 0
                                                \label{CR:pp}\\
    {[}\,\widehat g_{ab}(x),\widehat p^{cd}(x')\,] &=&
          i\hbar\,\delta^c_{(a}\delta^d_{b)}\,\delta^{(3)}(x,x')
                \label{CR:gp}
\end{eqnarray}
of operators defined on the $3$-manifold $\Sigma$.
Following Dirac, the constraints are interpreted as
constraints on the allowed state vectors $\Psi$, so that
$\widehat {\cal H}_a(x)\Psi=0=\widehat{\cal H}_\perp(x)$
for all $x\in\Sigma$. In particular, on choosing the states
as functions of the three-geometry $g$---and with operator
representatives $(\widehat
g_{ab}(x)\Psi)[g]:=g_{ab}(x)\Psi[g]$ and $(\widehat
p^{cd}(x)\Psi)[g]:=-i\hbar\delta\Psi[g]/\delta
g_{ab}(x)$---the constraints $\widehat{\cal H}_a\Psi=0$
imply that $\Psi[g]$ is constant under changes of $g$
induced by infinitesimal diffeomorphisms of the spatial
3-manifold $\Sigma$; and the crucial constraint $\widehat{\cal
H}_\perp(x)\Psi=0$ becomes the famous Wheeler-DeWitt
equation.

But the Wheeler-DeWitt equation is horribly ill-defined in any
exact mathematical sense and, unfortunately, perturbative
approaches to its definition and solution are as virulently
badly behaved as are its particle-physics based cousins: in
both cases the problem is trying to define products of
operator fields defined at the same point. Indeed, until
the rise of the Ashtekar programme (see below), most of the
work developing and using the Wheeler-DeWitt equation in
anything other than a totally formal sense relied on
truncating the gravitational field to just a few degrees of
freedom, so that it becomes a partial differential equation
in a finite number of variables, which one can at least
contemplate attempting to solve exactly.\footnote{So one
should not attach too much weight to the results of such
simple approximations. Admittedly, models of this type can
be valuable tools for exploring the many conceptual
problems that arise in quantum cosmology.}

In this context, we should mention again the Euclidean
programme in quantum gravity. Here, the central role is
played by functional integrals over all the
Riemannian---rather than Lorentzian---metrics on a
four-dimensional manifold $\cal M$. (The motivation for
Riemannian metrics is partly an analogy with the successful
use of imaginary time in Yang-Mills theory.) It can be
shown that the functional $\Psi[g]$ of $g$ defined by
certain such functional integrals satisfies (at least, in a
heuristic way) the Wheeler-DeWitt equation;
in particular, this is the basis of the famous
Hartle-Hawking `no boundary' proposal for the
`wave-function of the universe' in quantum cosmology. So in
this sense, the Euclidean programme amounts to a way of
constructing states for quantum geometrodynamics. For this
reason, and also because we discuss this programme (and its
use in quantum cosmology) in detail elsewhere \cite{BI99a},
we set it aside here.\footnote{A convenient recent source
for many of the original articles is Gibbons and Hawking
\cite{GH93}.}

\subsubsection{The Ashtekar programme}\label{SubSubSec:Ashtekar}
As emphasised above, the Wheeler-DeWitt equation is
ill-defined in any exact mathematical sense. However, a
major advance took place in 1986 when Ashtekar \cite{Ash86}
found a set of canonical variables which produce a dramatic
simplification of the structure of the central constraint
functions ${\cal H}_a(x)$ and ${\cal H}_\perp(x)$. Since
then there has been a very active programme to exploit
these new variables, both in classical general relativity
and in quantum gravity.

From a technical perspective, one of the great dangers in
canonical quantum gravity is the generation of anomalous
quantum excitations of non-physical modes of the
gravitational field. However, even to talk of such things
requires the operators to be defined rigorously---a task
that is highly non-trivial, since this is the point at
which the infamous ultraviolet divergences are likely to
appear. One of the main reasons why the Ashtekar programme
is potentially so important is the hope it offers of being
able to define these operators properly, and hence address
such crucial issues as the existence of anomalous
excitations.

The developments in the last decade have been very impressive
and, in particular, there is now real evidence in support of the
old idea that non-perturbative methods must play a key role in
constructing a theory of quantum gravity. If successful in its
current form, this programme will yield a theory of quantum
gravity in which unification of the forces is {\em not\/} a
necessary ingredient. Thus the Ashtekar programme demonstrates
the importance of distinguishing between a quantum theory of
gravity itself, and a `theory of everything' which of necessity
includes gravity.

In spite of their great structural significance, the use of
the Ashtekar variables has had little impact so far on the
conceptual problems in canonical quantum gravity, and so we
shall not discuss the technical foundations of this
programme here. However, it is important to note that one
of the new variables is a spin-connection, which suggested
the use of a gravitational analogue of the gauge-invariant
loop variables introduced by Wilson in Yang-Mills theory.
Seminal work in this area by Rovelli and Smolin \cite{RS90}
has produced many fascinating ideas, including a
demonstration that the area and volume of space are
quantised---something that is evidently of philosophical
interest and which has no analogue in quantum
geometrodynamics. We shall return to these ideas briefly in
Section 5.

\subsubsection{Spacetime According to the
Canonical Quantum Gravity Programme} The response given by
the canonical quantum gravity programme to the four probes
in Section \ref{SubSub:FocusQuestion} is broadly as
follows.

\smallskip\noindent
1. {\em Use of standard quantum theory\/}. The basic
technical ideas of standard quantum theory are employed,
suitably adapted to handle the non-linear constraints
satisfied by the canonical variables. On the other hand,
the traditional, Copenhagen interpretation of quantum
theory is certainly {\em not\/} applicable unless a
background spatial metric is assumed, and a resolution is
found of the problem of time, at least at some
semi-classical level. However, most attempts to implement
the canonical scheme abhor the introduction of any type of
background metric, and hence major conceptual problems can
be expected to arise if this programme is ever fully
realised.

Of course, one radical strategy for coping with this
situation is `to make a virtue of necessity', as discussed
in Section \ref{SubSubSec:ExtraValues}; {\em i.e.\/}, to
adopt the pilot-wave interpretation of quantum theory, and
thereby introduce a background metric in a strong sense,
involving a preferred foliation of spacetime. Note,
however, that with its emphasis on configuration space,
this interpretation would not seem appropriate for the
Ashtekar programme where, unlike quantum geometrodynamics,
the states are {\em not\/} functions on the configuration
space of all $3$-metrics.

Although we are setting aside quantum cosmology, we should
add that since most work in quantum cosmology has been done
within the canonical quantum gravity programme, there is
another sense in which the Copenhagen interpretation is
certainly not applicable to canonical quantum gravity;
whereas (as discussed in Section
\ref{SubSubSec:Literalism}), rivals such as Everettian
interpretations might be.

\smallskip\noindent
2. {\em Use of standard spacetime concepts\/}. In this
regard, canonical quantum gravity in effect `lies between'
the conservatism of the particle-physics programme, and the
radicalism of superstrings. For like the former, it uses a
background dimensional manifold (but it uses no background
metric). More precisely, the canonical theory of classical
relativity assumes {\em ab initio\/} that the spacetime
manifold $\cal M$ is diffeomorphic to $\Sigma\times\mathR$
where $\Sigma$ is some $3$-manifold; and this $3$-manifold
becomes part of the fixed background in the quantum
theory---so that, as for the particle-physics programme,
there is no immediate possibility of discussing quantum
changes in the spatial topology.

\smallskip\noindent
3. {\em The spacetime diffeomorphism group\/}. In the
canonical quantum gravity programme, the classical
Poisson-bracket algebra of the constraint functions ({\em
i.e.}, the Dirac algebra) can be interpreted as the algebra
of spacetime diffeomorphisms projected along, and normal
to, spacelike hypersurfaces. In the quantum theory, it is
usually assumed that this Poisson-bracket algebra is to be
replaced with the analogous commutator algebra of the
corresponding quantum operators

The Dirac algebra contains the group of spatial
diffeomorphisms, ${\rm Diff}(\Sigma)$, as a subgroup but it
is not itself a genuine group. Invariance under ${\rm
Diff\/}(\Sigma)$ means that the functionals of the
canonical variables that correspond to physical variables
are naturally construed as being non-local with respect to
$\Sigma$. Recent work with the loop-variable approach to
canonical quantum gravity has been particularly productive
in regard to the implications of invariance under spatial
diffeomorphisms.

The role of the full Dirac algebra is more subtle and
varies according to the precise canonical scheme that is
followed. There are still contentious issues in this
area---particularly in regard to exactly what counts as an
`observable' in the canonical scheme.

\smallskip\noindent
4. {\em The problem of time\/}. One of the main aspirations
of the canonical approach to quantum gravity has always
been to build a formalism with no background spatial, or
spacetime, metric; (this is particularly important, of
course, in the context of quantum cosmology). In the
absence of any such background structure, the problem of
time becomes a major issue.

There are various obvious manifestations of this. One is that the
Wheeler-DeWitt equation makes no apparent
reference to time, and yet this is to be regarded as the crucial
`dynamical' equation of the theory! Another manifestation
concerns the starting canonical commutation relations Eqs.\
(\ref{CR:gg}--\ref{CR:gp}). The vanishing of a commutator like
Eq.\ (\ref{CR:gg}) would normally reflect the fact that the
points $x$ and $x'$ are `spatially separated'. But what does this
mean in a theory with no background causal structure?

    The situation is usually understood to imply that, as
mentioned in Section \ref{SubSubSec:TimeGRProbTime}
(strategies 2 and 3), `time' has to be reintroduced as the
values of special {\em physical\/} entities in the
theory---either gravitational or material---with which the
values of other physical quantities are to be correlated.
Thus, rather than talking about clocks {\em measuring\/}
time---which suggests there is some external temporal
reference system---we think of time as being {\em
defined\/} by a clock, which in this case means part of the
overall system that is being quantised. Thus physical time
is introduced as a reading on a `physical' clock.

Unfortunately, it is a major unsolved problem whether (i)
this can be done at all in an exact way; and (ii) if so,
how the results of two different such choices compare with
each other, and how this is related to spatio-temporal
concepts. In fact, there are good reasons for thinking that
it is not possible to find any `exact' internal time, and
that the standard notion of time only applies in some
semi-classical limit of the theory. In this way time would
be an emergent or phenomenological concept, rather like
temperature or pressure in statistical physics. We discuss
this line of thought further in Section
\ref{SubSec:STEmerNQu} of \cite{BI99a}; here we just
emphasise that it is specifically about time, not
spacetime---and in that sense, not this paper's concern.
Section \ref{Sec:TowardsQuST} {\em will\/} discuss the idea
that spacetime, though not distinctively time, is
phenomenological.

\section{Towards Quantum Spacetime?}
\label{Sec:TowardsQuST}
\subsection{Introduction: Quantisation and Emergence}
In this Section, we turn to discuss some treatments of
spacetime that are in various ways more radical than those
given by the programmes in Section \ref{Sec:ResearchProg}.
We shall adopt the classification in Section
\ref{SubSubSec:FourApproachesQG} of four types of approach
to quantum gravity. Recall that they were:
\begin{enumerate}
\item[(1)] quantising general relativity;

\item[(2)] quantising a different classical theory, while
still having general relativity emerge as a low-energy
(large-distance) limit;

\item[(3)] having general relativity emerge as a low-energy limit of
a quantum theory that is not a quantisation of a classical
theory; and finally, and most radically,

\item[(4)] having both general relativity and quantum theory
emerge from a theory very different from both.
\end{enumerate}
Thus in Section \ref{SubSec:QuBelowMetric}, we will discuss
type (2); in Section \ref{SubSec:STFromNQu}, type (3); and
in Section \ref{SubSec:STEmerNQu}, type (4). But it will
help set these discussions in context, to take up two
topics as preliminaries: (i) the relation of the programmes
in Section \ref{Sec:ResearchProg} to this classification;
and (ii) the notion of emergence, and its relation to
quantisation, in general.

\subsubsection{Some Suggestions from the Three Programmes}
\label{SubSub:SuggestThreeProg} It is easy to place the
three programmes in Section \ref{Sec:ResearchProg} within the
above classification. We have seen that the particle-physics and
canonical programmes are examples of type (1), while the
superstrings programme is an example of type (2) (at least in its
perturbative version). But we should add two remarks to `sketch
in the landscape' of this classification.
\begin{enumerate}
\item All three programmes are {\em
similar\/} in that the main way they go beyond what we
called the common treatment of spacetime of our `ingredient
theories' (viz.\ as a 4-dimensional manifold with a
(classical) Lorentzian metric), is by quantizing a quantity
that is a standard type of physical variable within the
context of classical physics defined using the familiar
tools of differential geometry. For the particle-physics
and canonical programmes (Sections
\ref{SubSec:ParticleProg}, \ref{SubSec:CanWQG}), this is
the spatiotemporal metric $\gamma$, and the 3-metric $g$
respectively. In the perturbative superstrings programme,
the variable concerned is the function $X$ that maps a loop
into the target spacetime. However, in this Section, we
will discuss treatments that in some way or other `go
beyond' quantizing standard classical objects.

\item On the other hand, we should stress that the
two main {\em current\/} programmes discussed in Section
\ref{Sec:ResearchProg} make various radical suggestions
about spacetime: suggestions which are not reflected by
their being classified as types (1) and (2) in the schema
above. We already saw some of these suggestions in Section
\ref{Sec:ResearchProg}. In particular, we saw that the
superstrings programme requires spacetime to have a
dimension which is in general not four; and---more
strikingly---recent work on non-perturbative approaches
suggests that more than one manifold may contribute at the
Planck scale, or that models based on `non-commutative
geometry' may be appropriate. And for the canonical quantum
gravity programme, we mentioned the discrete spectra of the
spatial area and spatial volume quantities: results that
arguably suggest some type of underlying discrete structure
of space itself.
\end{enumerate}

But these programmes also make other such suggestions. For
example, we will see in Section \ref{SubSec:QuBelowMetric}
a general way in which quantising a metric (as in these
programmes) suggests a quantisation of logically weaker
structure such as differential or topological structure;
these are called `trickle-down effects'.

To sum up: Both these programmes threaten the ingredient
theories' common treatment of spacetime, quite apart from
their quantising the metric; indeed they even threaten the
manifold conception of space or spacetime.

\subsubsection{Emergence and Quantisation}
At this point it is worth developing the contrast
(introduced in Section \ref{SubSec:Prologue}) between these
two general strategies which one can adopt when attempting
to go beyond the common treatment of spacetime. The
distinction can be made in terms of any part of the common
treatment, not just metrical structure on which the
programmes in Section \ref{Sec:ResearchProg} focus: for
example, topological structure.

To `go beyond' such a structure, one strategy is to argue
that it is emergent (in physics' jargon:
`phenomenological'). `Emergence' is vague, and indeed
contentious; for along with related notions like reduction,
it is involved in disputes, central in philosophy of
science, about relations between theories and even
sciences. But here, we only need the general idea of one theory
$T_1$ being emergent from another $T_2$ if in a certain
part of $T_2$'s domain of application (in physics' jargon,
a `regime': usually specified by certain ranges of values
of certain of $T_2$'s quantities), the results of $T_2$ are
well approximated by those of $T_1$---where `results' can
include theoretical propositions as well as observational
ones, and even `larger structures' such as derivations and
explanations.

\begin{figure}[t]
\begin{picture}(400,330)(0,100)
    \put(0,400){\makebox(180,25){The `ultimate' theory\ \ \qquad
        $\longleftrightarrow$}
            \framebox(100,25){Exact results}}
    \put(230,370){$\Bigg\downarrow$\ \ emergence}
    \put(0,320){\makebox(180,25){Phenomenological theory\qquad
        $\longleftrightarrow$}
            \framebox(100,25){Limited results}}
    \put(230,290){$\Bigg\downarrow$\ \ emergence}
    \put(0,240){\makebox(180,25){Phenomenological theory\qquad
        $\longleftrightarrow$}
            \framebox(100,25){Limited results}}
    \multiput(233,185)(0,10){5}{\line(0,1){5}}
    \put(230,163){$\Bigg\downarrow$}
    \put(245,200){emergence}
\put(0,115){\makebox(175,25){General Relativity\qquad
$\longleftrightarrow$}
            \framebox(115,25){Limited results}}

\put(90,90){Figure 1.\ A hierarchy of phenomenological
theories}
\end{picture}

\end{figure}

This relation of emergence can of course be iterated,
yielding the idea of a `tower' of theories, each emerging
from the one above it. Figure 1 portrays this idea, for the
case of interest to us, viz.\ where the `bottom theory' is
classical general relativity. Figure 1 also uses the
physics jargon of a theory being `phenomenological'; and for the sake of
definiteness, it assumes an uppermost `ultimate'
theory---an assumption to which, as is clear from Section
\ref{SubSec:Realism?}, we are not committed. We should add
that of course many different towers will in general branch
off from a given theory.\footnote{For more discussion of
emergence, especially in relation to reduction, see Section
2 of our complementary essay \cite{BI99a}.}

So much for the general idea of emergence. The other
strategy for `going beyond a classical structure' is to try
to quantize it, in some sense; and then to recover it as
some sort of classical limit of the ensuing quantum theory.
We say `quantize in some sense', because although
`quantization' is certainly less vague than `emergence', it
is far from being precise. There is an open-ended family of
`quantization procedures' that are only provably equivalent
on some simple cases (such as certain finite-dimensional,
unconstrained Hamiltonian systems); and there is no
procedure known whereby an arbitrary classical system can
be quantised in an unequivocal way. Similarly, we say `some
sort of classical limit', because (as we said in Section
\ref{SubSubSec:FourApproachesQG}) `classical limit' also
has various senses: it can refer to special states whose
evolution over time follows classical laws, or to certain
quantum quantities taking values in a range where classical
theory is successful.

Though these two strategies are vague, the general idea of
them is enough to make it clear that they are independent.
That is: neither implies the other, though of course they
{\em can\/} be combined; (as the classification of
approaches (1)--(4) makes clear).

Thus one can maintain that some classical structure is
emergent, without quantizing it ({\em i.e.\/}, exhibiting
it as a classical limit of a quantisation). Indeed, there
are at least two ways one can do this. The first is
familiar from our discussion of superstrings (Section
\ref{SubSec:Superstrings}): a classical structure (there,
the metric geometry of general relativity) can emerge from
a quantum theory (superstrings) which is not a quantized
version of a classical theory of the structure. The second
way is independent of quantum theory: viz.\ a classical
structure could emerge from some theory which has nothing
to do with quantum theory. (In the context of quantum
gravity, this way is taken by approach (4) in the
classification of Section
\ref{SubSubSec:FourApproachesQG}.)

The converse implication can also be questioned. Quantizing
a structure does not {\em ipso facto\/} render the original
classical structure emergent. Agreed, if we quantize a
structure, then we can investigate the resulting quantum
theory's classical limits: and even allowing for the
vagueness of `quantisation' and `classical limit', we are
more or less guaranteed to be able to identify the original
classical structure as such a limit, or as a feature of
such a limit. But in view of the vagueness of `emergence',
this might not count as showing the classical structure to
be emergent within the quantum theory. Indeed, quite apart
from subtleties about the philosophical notion of
`emergence', the measurement problem of quantum theory
looms over the interpretation of such limits as `recovering
the classical world'.

So much by way of general discussion of these two
strategies, and their independence. The upshot, for our
topic of how to depart more radically from the common
treatment of spacetime than `just' by quantizing metrical
structure, is that one can apply either or both of these
two strategies to other classical structures---for example
to topological structure. We will see such applications in
more detail in the next three Subsections' discussions of
approaches (2)--(4) respectively.

Before embarking on those discussions, we should enter two
{\em caveats\/}. First: the ideas we are about to discuss
are far less established and understood, than are the trio
of {\em bona fide\/} research programmes in Section
\ref{Sec:ResearchProg}; so our discussion needs must be
much more tentative. Second: we will downplay the first
strategy, {\em i.e.\/}, emergence, on the grounds that our
complementary paper \cite{BI99a} discusses it---both in
general philosophical terms, and (for classical spacetime
structure), in terms of classical limits of a theory of
quantum gravity. Also {\em very\/} little is known about
the prospects for this strategy used on its own, {\em
i.e.\/}, uncombined with the second strategy of
quantisation---as we will see in Sections
\ref{SubSec:STFromNQu} and \ref{SubSec:STEmerNQu}.

\subsection{Quantisation `Below' the Metric}
\label{SubSec:QuBelowMetric} We turn to discuss approach
(2), where one quantises a theory other than classical
general relativity, but obtains it as a low-energy limit of
the quantised theory. Needless to say, only a tiny fraction
of the vast range of classical theories has been quantised,
so that in full generality, little is known about this
approach; and much of what is known concerns the
superstrings programme.

 So in this Subsection, we will confine ourselves to general
comments about the treatment of spacetime to be expected
once one adopts this approach. Furthermore, we will only
consider applying this approach to quantising classical
spacetime structures; (hence this Subsection's title).
Given this restriction, our discussion is given a natural
structure by the fact that the common treatment of
spacetime---viz.\ as a pair
$({\cal M},\gamma)$, with $\cal M$ a 4-dimensional
differentiable manifold, and $\gamma$ a Lorentzian
metric---appears at one end of a hierarchical chain of
structure; so that we can picture the introduction of a
quantum effect `below the metric $\gamma$' in terms of an
earlier point in the chain. Of course, a given mathematical
structure can often be placed in more than one such chain,
and the question then arises of which chain to use.

The common treatment of spacetime as a pair $({\cal
M},\gamma)$ fits naturally into the chain
\begin{equation}
\mbox{set of spacetime
points}\rightarrow\mbox{topology}\rightarrow
\mbox{differential structure}\rightarrow ({\cal M},\gamma)
                    \label{chain1}
\end{equation}
Indeed, this chain is implicit in much of our previous
discussion. The bottom level is a set $\cal M$, whose
elements are to be identified with spacetime points; but
this set is formless, its only general mathematical
property being its cardinal number. In particular, there
are no relations between the elements of $\cal M$, and no
special way of labelling any such element. The next step is
to impose a topology on $\cal M$ so that each point
acquires a family of neighbourhoods. Then one can talk
about relationships between points, albeit in a rather
non-physical way. This defect is overcome by adding the key
ingredient of the conventional treatment of spacetime: the
topology of $\cal M$ must be compatible with that of a
differentiable manifold, so that a point in $\cal M$ can be
labelled uniquely (at least, locally) by giving the values
of four real numbers. In the final step a Lorentzian metric
$\gamma$ is placed on $\cal M$, thereby introducing the
ideas of the lengths of a path joining two spacetime
points, parallel transport with respect to a Riemannian
connection, causal relations between pairs of points etc.
Note that an analogous discussion applies to the usual
modelling of space and time individually by a
$3$-dimensional, and $1$-dimensional manifold respectively.

Note that a variety of intermediate stages can be inserted:
for example, the link `$\mbox{differential
structure}\rightarrow({\cal M},\gamma)$' could be factored
as
\begin{equation}
\mbox{differential structure}\rightarrow
    \mbox{causal structure}\rightarrow({\cal M},\gamma).
\end{equation}

    A quite different scheme arises by exploiting the fact that a
differentiable manifold $\cal M$ is uniquely determined by
the algebraic structure of its commutative ring of
differentiable functions, ${\cal F}({\cal M})$ (cf.\ our
discussion in Section \ref{SubSec:InterpretGR} of the
spectral theorem for commutative C*algebras.) And a ring is
itself a complicated algebraic structure that can be
analysed into a hierarchy of substructures in several ways.
Thus one alternative chain to Eq.\ (\ref{chain1}) is
\begin{equation}
\mbox{set}\rightarrow\mbox{abelian group}\rightarrow
\mbox{vector space}\rightarrow{\cal F}({\cal M})\rightarrow
    ({\cal M},\gamma).              \label{chain2}
\end{equation}

Given these chains of structure leading to $({\cal
M},\gamma)$, and others like them, it is clear what are the
options facing approach (2). One has to decide:
\begin{enumerate}
\item[(i)] which of the chains to $({\cal M},\gamma)$ to use;
and
\item[(ii)] at what level in the chosen chain to try to
quantise.
\end{enumerate}
For example, if one uses the first chain, one faces such
questions as: should we accept a fixed set of spacetime
(or, for a canonical approach: spatial) points, but let the
topology and/or differential structure be subject to
quantum effects? Or should we say that the notion of a
spacetime point itself is not meaningful at a fundamental
level: {\em i.e.}, it is a concept that should not appear
in the theory, even in a mathematical sense?

We end by making two general comments about this situation,
corresponding to the two decisions, (i) and (ii).

In regard to (i), we stress that the details of one's
programme will depend strongly on the initial decision
about which chain to use. Thus if one decides to apply
quantization to the second chain Eq.\ (\ref{chain2}), one
is led naturally to consideration of the algebraic approach
to classical general relativity pioneered by Geroch
\cite{Ger72} (`Einstein algebras') and non-commutative
analogues thereof \cite{PZ95}. And, of course, the idea of
a non-commutative version of the algebra ${\cal F}({\cal
M})$ was one of the motivating factors behind Connes'
seminal ideas on non-commutative geometry \cite{Con94}.

 Concerning (ii), one should be aware that, once a chain
has been chosen, quantisation at one level could `trickle
down' to produce quantum effects at a `lower', more
general, level in the chain; ({\em i.e.\/} to the left in
our diagrams). For example, quantisation of the metric
could trickle down to the differential structure or
topology.\footnote{This was mentioned in Section
\ref{SubSub:SuggestThreeProg}, as a way in which the
programmes in Section \ref{Sec:ResearchProg} put pressure
on the manifold conception of spacetime; a way additional
to those already discussed.}

 Such `trickle-down' effects were   envisaged by Wheeler \cite{Whe68} in
his original ideas about quantum topology in the context of
canonical quantization. His idea was that large quantum
fluctuations in a quantized 3-metric $\hat{g}_{ab}(x)$
would generate changes in the spatial topology; for the
effects of quantum gravity would become more pronounced at
decreasing distances, resulting eventually in a `foam like'
structure at around the Planck length.

We stress that such effects depend on an appropriate
mechanism. Thus, {\em contra\/} Wheeler's intuition that
quantum gravity effects become stronger at decreasing
distances, one might hold that, in fact, quantum gravity is
`asymptotically-free'---so that the effects become {\em
smaller\/} as the scale reduces. Under these circumstances,
there would be no metric-driven topology
changes.\footnote{The idea that gravity might be
asymptotically-free was studied some years ago by Fradkin
and Tseytlin \cite{ET81} in the context of $R+R^2$ theories
of gravity.}

Another, more radical, example of trickle-down effects
arises in connection with Penrose's thesis that a
projective view of spacetime structure is more appropriate
in quantum gravity: in particular, a spacetime point should
be identified with the collection of all null rays that
pass through it. Quantising the spacetime metric will then
induce quantum fluctuations in the null rays, which will
therefore no longer intersect in a single point. In this
way, quantum fluctuations at the top of the first chain
Eq.\ (\ref{chain1}) trickle right down to the bottom of the
chain, so that the very notion of a `spacetime point'
acquires quantum overtones.

\subsection{Spacetime from a Non-Quantisation}
\label{SubSec:STFromNQu} We turn to approach (3) in our
classification, according to which general relativity emerges as
a low-energy limit of a quantum theory that is {\em not\/} given
as the quantisation of a classical theory, but rather
`intrinsically' in some way.

By and large, programmes following this approach will
reject the conception of spacetime as a manifold (`from the
outset', rather than `sneaking up' on this conclusion in
the way the programmes in Section \ref{Sec:ResearchProg}
do). Agreed, the envisaged quantum theory might in
principle include the postulation of a spacetime manifold
(though it is not a quantisation of a classical theory
defined on that manifold): a manifold which then turns out
to be the manifold on which the low-energy limit, general
relativity, is defined. But by and large, this circumstance
would be odd: why should the quantum theory postulate just
what the emergent approximation needs? Accordingly, we will
here consider programmes (or, more precisely, mainly one
fragment of one possible programme!) which do indeed reject
the spacetime manifold at the fundamental level.

An example of the way the envisaged quantum theory might
have general relativity as a low-energy limit is as
follows. Return to the tower of theories of Section
\ref{SubSec:QuBelowMetric}, and imagine that one result of
emergence at some level in the tower is: (i) the idea of a
`local region'---not regarded as a subset of something
called `spacetime', but rather as an emergent concept in
its own right; together with (ii) an algebra of such
regions that specifies their theoretical use, and that can
be identified mathematically as the algebra of a certain
open covering of a genuine continuum manifold $\cal M$.
Hence---as long as one keeps to the phenomena appropriate
to this level---it is {\em as if\/} physics is based on the
spacetime manifold $\cal M$, even though this plays no
fundamental role in the `ultimate' theory with which we
started. Specific ideas of this type have arisen in the
context of attempts to quantise the point-set topology of a
set \cite{Sor91a,Ish90d}. We note in passing that the
mathematics of locales, and more generally topos theory (in
particular, the idea of a `Grothendieck topos') provides a
natural framework in which to develop the idea that regions
are more important than points.

    The possible significance of regions, rather than points,
arises also in recent ideas about the nature of quantum physics
in a bounded region. These go back to an old remark of Bekenstein
\cite{Bek74} to the effect that any attempt to place a quantity
of energy $E$ in a spatial region with boundary area $A$---and
such that $E>\sqrt A$---will cause a black hole to form, and this
puts a natural upper bound on the value of the energy in the
region. The implication is that in any theory of quantum gravity
whose semi-classical states contain something like black-hole
backgrounds, the quantum physics of a bounded region will involve
only a {\em finite\/}-dimensional Hilbert space. This intriguing
possibility is closely related to the so-called `holographic'
hypothesis of t'Hooft \cite{tHoo93} and Susskind \cite{Sus95} to
the effect that physical states in a bounded region are described
by a quantum field theory on the {\em surface\/} of the region,
with a Hilbert space of states that has a finite dimension.

    Ideas of this type could have major implications for
quantum gravity. In particular---and in terms of the tower
in Figure 1---the implication is that at one level of
phenomenological theory the idea of local spacetime regions
makes sense, and in those regions the quantum theory of
gravity is finite-dimensional. However, in the---possibly
different---tower of phenomenological approximations that
includes weak-field perturbative approaches to quantum
gravity, the effective theory uses an infinite-dimensional
Hilbert space to describe the states of weakly-excited
gravitons. And, at this level, spacetime is modelled by a
continuum manifold, with a full complement of spacetime
points.

We should emphasise that an approach like this does not
necessarily exclude the proposals of the more familiar
research programmes in quantum gravity: such proposals
would however become phenomenological, {\em i.e.\/}, part
of the emergence of general relativity and quantum theory.
If our present understanding of quantum gravity is any
guide, this effective quantisation of the gravitational
field will involve a non-local---possibly
string-like---structure. This raises the intriguing
question of whether superstring theory and the
loop-variable approach to canonical quantum gravity can
both be regarded as different modes---or phases---of a more
basic, common structure \cite{Smo98}. A central issue,
presumably, is whether supersymmetry can be assigned some
significant role in the Ashtekar programme.

\subsection{Spacetime emergent from a Non-Quantum Theory}
\label{SubSec:STEmerNQu}

Finally let us raise the question of the justification in
quantum gravity of the use of standard quantum theory
itself. So, in terms of Figure 1's portrayal of the
emergence of general relativity, the idea now is that there
would also be a tower of theories leading down to the
emergence of standard quantum theory. Of course, in
accordance with the point that towers can branch off from each
other, this tower may well not be the same one as that
leading to general relativity.

One particularly relevant issue in regard to quantum
theory---and the only one that we shall discuss in any
detail---is the question of what justifies its use of {\em
continuum\/} concepts: specifically, its use of real and
complex numbers. This question is very pertinent if one is
already worried about the use of continuum
ideas in the manifold model of space or time.

The formalism of quantum theory immediately suggests two
answers: one concerning eigenvalues, and the other
probabilities. Thus one might answer by saying that real
numbers represent the possible results of measurements; (so
that if eigenvalues of operators are to represent results,
we want the operators to be self-adjoint). But why should
measurement results be represented by real numbers? One
natural, if not compelling, answer is that apparently all
measurement results can in principle be reduced to the
positions of a pointer in space---and space is modelled
using real numbers. At the very least, this is certainly
true of the elementary wave mechanics of a point particle
moving in standard space; and this example, particularly
the Hilbert space generalisation of its specific
mathematical structure---has become one of the paradigms
for quantum theory in general.

 So according to this line of thought, the use of real numbers (and
similarly: complex numbers) in quantum theory in effect
involves a prior assumption that space should be modelled
as a continuum. If so, then the suggestion that standard
spacetime concepts break down at the scale of the Planck
length and time, and must be replaced by some discrete
structure which only `looks like' a differentiable manifold
at larger scales, means that we cannot expect to construct
a theory of this discrete structure using standard quantum
theory---with its real and complex numbers. Of course, this
argument is not water-tight; but it does illustrate how
potentially unwarranted assumptions can enter speculative
theoretical physics, and thereby undermine the enterprise.

The second possible answer to the question `what justifies
quantum theory's use of real and complex numbers?' is that
probabilities are real numbers between 0 and 1; (so that if
probabilities are to be given by the squared norm of state
vectors, the vector space must have $\mathR$ or $\mathC$ as
its ground-field). But why should probabilities be
represented by real numbers? Of course, if probability is
construed as `relative frequencies' of sequences of
measurements, then the real numbers do arise naturally as
the ideal limits of collections of rational numbers.
However, the idea of measurement `at the Planck length' is
distinctly problematic (cf.\ our remarks in Section
\ref{SubSubSec:TranscendIdealism} about `in principle
inaccessibility'), and if the concept of probability is
relevant at all in such regimes, one may feel that a
different interpretation of this concept is more
appropriate: for example, the propensity interpretation.

But there is no {\em a priori\/} reason why a `propensity'
(whatever that may mean!) should be modelled by a real
number lying between $0$ and $1$. Agreed, it may well be
appropriate sometimes to say that one propensity is
`larger' than another; but there may also be propensities
that cannot be compared at all (a not unreasonable
suggestion in the context of non-commuting operators in a
quantum theory), and this suggests that a minimal model for
such probabilities would be a partially-ordered set with
some type of additional algebraic structure (so that `sums'
of probabilities can be defined for disjoint propositions).

For these reasons, a good case can be made that a
complete theory of quantum gravity may require a
revision of quantum theory itself in a way that removes
the {\em a priori\/} use of continuum numbers in its
mathematical formalism.

Finally, we note that, from time to time, a few hardy souls
have suggested that a full theory of quantum gravity may
require changing the foundations of mathematics itself. A
typical argument is that standard mathematics is based on
set theory, and certain aspects of the latter (for example,
the notion of the continuum) are grounded ultimately in our
spatial perceptions. However, our perceptions probe only
the world of classical physics---and hence we feed into the
mathematical structures currently used in {\em all\/}
domains of physics, ideas that are essentially classical in
nature. The ensuing category error can be remedied only by
thinking quantum theoretically from the very outset---in
other words, we must look for `quantum analogues' of the
categories of standard mathematics.

    How this might be done is by no means obvious.\footnote{A
recent example of this type of thinking can be found in a
book by Finkelstein\cite{Fink96}.} One approach is to claim
that, since classical logic and set theory are so closely
linked (a proposition $P$ determines---and is determined
by---the class of all entities for which $P$ can be rightly
asserted), one should start instead with the formal
structure of {\em quantum\/} logic and try to derive an
analogous `non-Boolean set theory'. Such ideas are related
to the exciting subject of topos theory, which can be
viewed as a far-reaching generalisation of standard set
theory. This is why, as mentioned in Section
\ref{SubSec:STFromNQu}, topos theory is a natural arena
within which to develop speculative schemes in which
`regions' of spacetime (or space, or time) are more
important than `points' (which may not exist at
all).\footnote{Topos theory has a related deep connection
with non-standard logical structures: something we have
exploited in recent work useing presheaf logic to analyse
the Kochen-Specker theorem in standard quantum theory
\cite{IB98a,BI99b}.}

\subsection{Envoi}
Clearly, this Section has opened up a Pandora's box of
possibilities for the overall shape of a theory of quantum
gravity: possibilities that it is well-nigh impossible to
adjudicate between---not least because it is very hard even
to make individual possibilities precise and detailed. So
we will make no pretence of judging them here.

Instead, by way of a modest and (mercifully brief!)
conclusion, we want to stress some of this paper's lines of
thought, that led to this Pandora's box. We began with the
question how spacetime might be treated in a theory of
quantum gravity. We described how the search for such a
theory was beset by various conceptual difficulties,
including difficulties about the ingredient theories, and
about spacetime concepts---and also beset by a dire lack of
data; (a predicament reviewed in Sections \ref{Sec:Intro}
to \ref{Sec:WhatisQG}). On the other hand, we could not
`duck out' of searching for some such theory (Sections
\ref{SubSubSec:MotivationsQG}--\ref{SubSubSec:QGAvoided?}).
We reviewed in Section \ref{Sec:ResearchProg} three main
programmes that all proceed by quantising a classical
theory which postulates a spacetime manifold. In various
ways these programmes suggest that there are fundamental
limitations to the applicability of the manifold conception
of spacetime; (of course, they also have various more
specific problems, both physical and conceptual). Thus our
attention was turned to the more radical programmes of this
Section \ldots where we admit to having to suspend
judgment.

To sum up: Quantum gravity is most unusual in comparison
with other branches of physics, and indeed with most other
branches of human enquiry---or with other `games people play'. It
is an exciting unpredictable game, with very few
rules---and yet, as the sports commentators say, `there is
everything to play for!'

\section*{Acknowledgements}
Chris Isham would like to thank the Mrs.\ L.~D.~Rope Third
Charitable Settlement for financial support.


\begin{thebibliography}{10}

\bibitem{BI99a}
J.~Butterfield and C.J.~Isham.
\newblock On the emergence of time in quantum gravity.
\newblock In J.~Butterfield, editor, {\em The Arguments of Time}. Oxford
  University Press, Oxford, 1999.
\newblock gr-qc/9901024.

\bibitem{Kai92}
D.~Kaiser.
\newblock More Roots of Complementarity: Kantian Aspects and Influences.
\newblock {\em Studies in History and Philosophy of Science}, 23:213--239, 1992.

\bibitem{Butt95}
J.~Butterfield.
\newblock Worlds, minds and quanta.
\newblock {\em Aristotelian Society Supplementary Volume}, 69:113--158, 1995.

\bibitem{Butt96}
J.~Butterfield.
\newblock Whither the minds?
\newblock {\em British Journal for the Philosophy of Science}, 47:200--221,
  1996.

\bibitem{Rid99b}
K.~Ridderbos.
\newblock The loss of coherence in quantum cosmology.
\newblock {\em Studies in History and Philosophy of Modern Physics},
30B:41--60, 1999.

\bibitem{Har95}
J.~Hartle.
\newblock Spacetime quantum mechanics and the quantum mechanics of spacetime.
\newblock In B.~Julia and J.~Zinn-Justin, editors, {\em Proceedings of the 1992
  Les Houches School, Gravitation and Quantisation}, pages 285--480. Elsevier
  Science, 1995.

\bibitem{IL94}
C.J.~Isham and N.~Linden.
\newblock Quantum temporal logic and decoherence functionals in the histories
  approach to generalised quantum theory.
\newblock {\em Journal of Mathematical Physics}, 35:5452--5476, 1994.

\bibitem{Val96}
A.~Valentini.
\newblock {\em On the Pilot-Wave Theory of Classical, Quantum and Subquantum
  Physics}.
\newblock Springer-Verlag, Berlin, 2000.

\bibitem{Bub97}
J.~Bub.
\newblock {\em Interpreting the Quantum World}.
\newblock Cambridge University Press, Cambridge, 1997.

\bibitem{GRW86}
G.C.~Ghirardi, A.~Rimini, and T.~Weber.
\newblock Unified dynamics for microscopic and macroscopic systems.
\newblock {\em Physical Review}, D34:470--491, 1986.

\bibitem{Pea89}
P.~Pearle.
\newblock Combining stochastical dynamical state-vector reduction with
  spontaneous localization.
\newblock {\em Physical Review}, A39:2277--2289, 1989.

\bibitem{PS95}
P.~Pearle and E.~Squires.
\newblock Gravity, energy conservation and parameter values in collapse models.
\newblock 1995.
\newblock quant-ph/9503019.

\bibitem{Ear95}
J.~Earman.
\newblock {\em Bangs, Crunches, Whimpers and Shrieks}.
\newblock Oxford University Press, Oxford, 1995.

\bibitem{EarNor87}
J.~Earman and J.~Norton.
\newblock What price space-time substantivalism? The hole story.
\newblock {\em British Journal for the Philosophy of Science}, 38:515--525,
  1987.

\bibitem{Ear89}
J.~Earman.
\newblock {\em World Enough and Space-Time}.
\newblock MIT Press, Cambridge, Massachusetts, 1989.

\bibitem{Jac95}
T.~Jacobson.
\newblock Thermodynamics of spacetime: {T}he {E}instein equation of state.
\newblock {\em Physical Review Letters}, 75:1260, 1995.

\bibitem{BR33}
N.~Bohr and L.~Rosenfeld.
\newblock Zur frage der messbarkeit der elektromagnetischen feldgrossen.
\newblock {\em Kgl.\ Danek Vidensk.\ Selsk.\ Math.-fys.\ Medd.}, 12:8, 1933.

\bibitem{Ros63}
L.~Rosenfeld.
\newblock On quantization of fields.
\newblock {\em Nuclear Physics}, 40:353--356, 1963.

\bibitem{PG81}
D.N.~Page and C.D.~Geilker.
\newblock Indirect evidence for quantum gravity.
\newblock {\em Physical Review Letters}, 47:979--982, 1981.

\bibitem{Fey63}
R.~Feynman.
\newblock Lectures on gravitation.
\newblock {\em Acta Physical Polonica}, XXIV:697, 1963.

\bibitem{BD75}
D.G.~Boulware and S.~Deser.
\newblock Classical general relativity derived from quantum gravity.
\newblock {\em Annals of Physics}, 89:193, 1975.

\bibitem{Don98}
J.~Donoghue.
\newblock Perturbative dynamics of quantum general relativity.
\newblock In {\em Proceedings of the Eighth Marcel Grossmann Conference on
  General Relativity}. 1998.

\bibitem{Kuc92a}
K.~Kucha\v{r}.
\newblock Time and interpretations of quantum gravity.
\newblock In {\em Proceedings of the 4th Canadian Conference on General
  Relativity and Relativistic Astrophysics}, pages 211--314. World Scientific,
  Singapore, 1992.

\bibitem{Ish93}
C.J. Isham.
\newblock Canonical quantum gravity and the problem of time.
\newblock In {\em Integrable Systems, Quantum Groups, and Quantum Field
  Theories}, pages 157--288. Kluwer Academic Publishers, London, 1993.

\bibitem{UW89}
W.~Unruh and R.M. Wald.
\newblock Time and the interpretation of quantum gravity.
\newblock {\em Physical Review}, D40:2598--2614, 1989.

\bibitem{IPS75}
C.J. Isham, R.~Penrose, and D.W.~Sciama.
\newblock {\em Quantum Gravity: {A}n {O}xford Symposium}.
\newblock Clarendon Press, 1975.

\bibitem{IPS81}
C.J. Isham, R.~Penrose, and D.W.~Sciama.
\newblock {\em Quantum Gravity: {A} Second {O}xford Symposium}.
\newblock Clarendon Press, 1981.

\bibitem{Ish92}
C.J. Isham.
\newblock Conceptual and geometrical problems in quantum gravity.
\newblock In H.~Mitter and H.~Gausterer, editors, {\em Recent Aspects of
  Quantum Fields}, pages 123--230. Springer-Verlag, Berlin, 1992.

\bibitem{Kuc93}
K.~Kucha\v{r}.
\newblock Canonical quantum gravity.
\newblock In R.J. Gleiser, C.N. Kozameh, and O.M. Moreschi, editors, {\em
  General Relativity and Gravitation, 1992}, pages 119--150. IOP Publishing,
  Bristol, 1993.

\bibitem{GH93}
M.~Gell-{M}ann and J.~Hartle.
\newblock Classical equations for quantum systems.
\newblock {\em Physical Review}, D47:3345, 1993.

\bibitem{Ash86}
A.~Ashtekar.
\newblock New variables for classical and quantum gravity.
\newblock {\em Physical Review Letters}, 57:2244--2247, 1986.

\bibitem{RS90}
C.~Rovelli and L.~Smolin.
\newblock Loop space representation of quantum general relativity.
\newblock {\em Nuclear Physics}, B331:80--152, 1990.

\bibitem{Ger72}
R.~Geroch.
\newblock Einstein algebras.
\newblock {\em Communications in Mathematical Physics}, 25:271--275, 1972.

\bibitem{PZ95}
G.N. Parfionov and R.R. Zapatrin.
\newblock Pointless spaces in general relativity.
\newblock {\em International Journal of Theoretical Physics}
            34:717, 1995.

\bibitem{Con94}
A.~Connes.
\newblock {\em Non Commutative Geometry}.
\newblock Academic Press, New York, 1994.

\bibitem{Whe68}
J.A. Wheeler.
\newblock Superspace and the nature of quantum geometrodynamics.
\newblock In C.~DeWitt and J.W. Wheeler, editors, {\em Battelle Rencontres:
  1967 Lectures in Mathematics and Physics}, pages 242--307. Benjamin, New
  York, 1968.

\bibitem{ET81}
E.S. Fradkin and A.A. Tseytlin.
\newblock Renormalizable asymptotically free quantum theory of gravity.
\newblock {\em Physics Letters}, 104B:377--381, 1981.

\bibitem{Sor91a}
R.D. Sorkin.
\newblock Finitary substitute for continuous topology.
\newblock {\em International Journal of Theoretical Physics}, 30:923--947, 1991.

\bibitem{Ish90d}
C.J. Isham.
\newblock An introduction to general topology and quantum topology.
\newblock In H.C. Lee, editor, {\em Physics, Geometry and Topology}, pages
  129--190. Plenum Press, New York, 1990.

\bibitem{Bek74}
J.D. Bekenstein.
\newblock The quantum mass spectrum of a {K}err black hole.
\newblock {\em Lettere a Nuovo Cimento}, 11:467--470, 1974.

\bibitem{tHoo93}
G.~t'Hooft.
\newblock Dimensional reduction in quantum gravity.
\newblock 1993.
\newblock gr-qc/9310026.

\bibitem{Sus95}
L~Susskind.
\newblock The world as a hologram.
\newblock {\em Journal of Mathematical Physics},
36:6377--6396, 1995.

\bibitem{Smo98}
L.~Smolin.
\newblock Strings as perturbations of evolving
spin-networks.
\newblock hep-th/9801022.

\bibitem{Fink96}
D.R. Finkelstein.
\newblock {\em Quantum Relativity: {A} Synthesis of the Ideas of {E}instein and
  {H}eisenberg}.
\newblock Springer-Verlag, New York, 1996.

\bibitem{IB98a}
C.J. Isham and J.~Butterfield.
\newblock A topos perspective on the {K}ochen-{S}pecker theorem: {I.} {Q}uantum
  states as generalised valuations.
\newblock {\em International Journal of Theoretical Physics},
37:2669--2733, 1998.
\newblock quant-ph/980355.

\bibitem{BI99b}
J.~Butterfield and C.J. Isham.
\newblock A topos perspective on the {K}ochen-{S}pecker theorem: {II.}
  {C}onceptual aspects, and classical analogues.
\newblock {\em International Journal of Theoretical Physics}, 1999.
\newblock To appear.
\newblock quant-ph/9808067.

\end{thebibliography}
\end{document}